\documentstyle[prl,preprint,aps,psfig]{revtex}
\begin{document}
\draft
\title{Composite Fermions in the Hilbert Space of the
Lowest Electronic Landau Level}
\author{J.K. Jain and R.K. Kamilla}
\address{Department of Physics, State University of New York
at Stony Brook, Stony Brook, New York 11794-3800\\
Department of Physics, MIT, Cambridge, Massachussetts 02139}
\date{\today}
\maketitle
\begin{abstract}

Single particle basis functions for composite fermions are
obtained from which many-composite fermion states confined
to the lowest electronic Landau level can be constructed in
the standard manner, i.e., by building Slater determinants.
This representation  enables a Monte Carlo study
of systems containing a large number
of composite fermions, yielding new quantitative and qualitative
information. The ground state energy and the gaps
to charged and neutral excitations
are computed for a number of fractional quantum Hall
effect (FQHE) states, earlier off-limits to a quantitative
investigation. The ground state energies are
estimated to be accurate to $\sim$ 0.1\% and the gaps
at the level of a few percent.  It is also shown that
at Landau level fillings smaller than or equal to 1/9 the
FQHE is unstable to a spontaneous creation of excitons of
composite fermions. In addition, this approach
provides new conceptual insight
into the structure of the composite fermion wave functions,
resolving in the affirmative the question of whether it is
possible to motivate the composite fermion theory
entirely within the lowest Landau level, without appealing to
higher Landau levels.

\end{abstract}


\section{Introduction}

Interacting electrons confined to two dimensions and 
subjected to a strong magnetic field exhibit spectacular 
phenomena, e.g., the fractional
quantum Hall effect (FQHE) \cite {book}. 
An elegant and succinct qualitative explanation 
of these phenomena is given in terms of new particles 
called composite fermions 
\cite {Jain}, which are electrons bound to an even 
number of vortices of the many
body wave function. The reason for the simplicity of the 
composite fermion (CF) approach is that the 
strongly correlated liquid of electrons is mapped into a 
weakly interacting gas of composite fermions, enabling a 
conventional single-particle description of the FQHE, similar 
in spirit to that in Landau's theory of Fermi liquids.
The composite fermions have the same charge, spin and statistics 
as electrons, but they differ from the electrons in the 
fundamental non-perturbative aspect that they experience a 
reduced effective magnetic field, given by 
\begin{equation}
B^{CF}=B-2m\rho\phi_0
\end{equation}
where $B$ is the external magnetic field, $\rho$ is the 
electron (CF) density, and $\phi_0=hc/e$ is the flux quantum. 
This is a consequence of a partial cancellation of the 
Aharonov-Bohm phases by the phases generated by the vortices 
on other composite fermions.
The effective filling factor of composite fermions, 
$\nu^{CF}=\rho \phi_0 /|B^{CF}| $ is therefore given by 
\begin{equation}
\nu=\frac{\nu^{CF}}{2m\nu^{CF}\pm 1}\;\;,
\end{equation}
where $\nu=\rho \phi_0/B $ is the filling factor of electrons.
The minus sign corresponds to the case in which $B^{CF}$ and 
$B$ are antiparallel.

The model of weakly interacting composite fermions has 
explained numerous facts  pertaining to the 
FQHE \cite {book}, some of which are as follows. 
The fractions appear in experiments in certain prominent 
sequences, given by 
\begin{equation} \nu =\frac{n}{2pn\pm 1}\;\;,
\end{equation} 
which are called the principal sequences of fractions. These 
correspond in the CF theory to filled Landau levels (LLs) of composite 
fermions ($\nu^{CF}=n$), i.e., the FQHE of electrons is thus simply 
a manifestation of the IQHE of the composite fermions. 
The absence of other fractions 
is related to the scarcity of FQHE in 
higher LLs. The experimentally determined gaps to charged 
excitations also depend on 
$n$ roughly in a way expected from modeling them as the 
cyclotron energy of the composite fermions \cite {gaps}. 
An interesting limiting case is at $\nu=1/2$, where $B^{CF}$ 
vanishes. Halperin, Lee and Read have proposed that the composite 
fermions exist here as well and form a Fermi sea \cite {HLR}, which
has given rise to numerous new experiments \cite {semi} that 
detect composite fermions by observing  their 
cyclotron orbits in the vicinity of half filled 
Landau level and provide further nontrivial confirmation of the 
CF hypothesis. The CF approach can straightforwardly be 
extended to include spin as well \cite {Jain,Wuspin}, which is of 
interest since the Zeeman energy is quite 
small in GaAs. Du {\em et al.} \cite {spin}
have studied in detail the appearance and disappearance
of the fractions as a function of the Zeeman energy in terms of
transitions between incompressible states of various polarizations,
originating from a competition between the cyclotron and the 
Zeeman energies of the composite fermions. Thus, 
the CF theory appears to give a simple and unified description of 
all the {\em liquid} states of electrons in the lowest LL, 
incompressible or otherwise.

Equally striking is the success of 
the microscopic CF theory. The CF wave functions have
been compared to the exact wave functions (obtained for a 
finite number of electrons,
usually fewer than 10-12, by a brute force numerical 
diagonalization of the Hamiltonian in the lowest LL subspace) 
in a large number of studies and found to  
have an overlap of essentially 100\% \cite {Wuspin,Dev,RR}, which is 
particularly impressive considering 
that the CF wave functions have no adjustable parameters for 
the incompressible FQHE states and their low-energy 
excitations. [In contrast, in studies of {\em atoms}
involving Jastrow factors,
which typically depend on several parameters, typically only
80 to 85\% of the correlation energy is obtained correctly.
With backflow corrections and by including generalized Jastrow factors
that account for three-body interactions, the accuracy can be
increased to within a few \%, but only at the cost of introducing
many more adjustable parameters (see, for example Ref. \cite {Jastrow}).]
Therefore, it should in principle 
be straightforward to make accurate 
quantitative predictions for the experimentally measurable 
quantities using the CF wave functions.

Unfortunately, in the past, it has not been possible to 
compute with the CF wave functions for systems containing
more than $\sim $ 10 composite fermions due to the 
following technical complication.
In  the simplest (unprojected)
form, the CF wave function have some amplitude in higher electronic
LLs. Even though it turns out to be surprisingly small 
\cite {Trivedi}, the unprojected wave functions are not  
very useful for quantitative estimates, especially in the limit 
of large $B$. Here, a good quantitative account is given by the
lowest-LL-projected CF wave functions.
(Here and below, it will be understood that the 
phrase ``lowest Landau level" refers to
the lowest LL of electrons; composite fermions {\em themselves}
will be allowed to occupy any number of {\em CF}-LLs, which all
reside completely within the lowest electronic LL.)
In our earlier work, 
the projection was computed numerically (exactly) by expanding
the CF wave function into electronic basis states and throwing away the 
part containing higher LLs \cite {Dev}. However, the time and 
memory requirements increase exponentially for this method, 
limiting the projection to approximately 10 particles. 
It is also possible to express the projection operator in an 
integral form, but  Monte Carlo with these wave functions runs 
into the technical difficulty of the fermion sign problem.

As a result, our quantitative understanding  
of the FQHE in the past has been largely 
limited to what we can learn from the studies of small systems, 
either from exact diagonalization or using the CF wave functions
(both of which produce virtually identical results). The size of 
the many-body Hilbert space is the smallest closest to half filled
LL, and grows quickly away from $\nu=1/2$. As a result, the 
maximum number of electrons for which exact diagonalization
has been performed depends on the filling factor; 
at $\nu=$ 1/5, 1/3, 2/5 or 3/7,  it is $N_{max}=$ 
7, 8, 10, or 12 \cite {He,Fano}
and the largest system studied to date is 14 \cite {RR}. 
While systems containing 10-12 particles  are sufficiently large
for convincing ourselves of the validity of the
concept of composite fermions and testing
the accuracy of the CF wave functions,
they are pitifully small for making reliable
thermodynamic estimates for several quantities of interest. In fact,
several of the FQHE states do not even show up in such small
systems, which can be understood as follows.
In the conventional spherical geometry,  
the $n/(2n+1)$ FQHE state requires at least $n^2$ electrons, 
as seen by analogy to the IQHE: $n^2$  is the minimum 
number of fermions needed to fill $n$ LLs. As a result, only 
two data points are available for 3/7 ($N=9$ and 12), while 4/9,
5/11, {\em etc}. are totally inaccessible.
The situation is worse for the states at $\nu=n/(4n\pm 1)$ -- 
here, it is difficult to study even 2/9.
Due to an exponential increase with the number of electrons in the 
computer time and memory requirements, it is
safe to assume that at best marginal improvements will be possible
in the future exact diagonalization studies, which
will therefore not tell us much more 
than what they already have.  Consquently, our best hope 
for a better quantitative description of the FQHE lies
in developing new techniques to work with the CF wave functions.  

Progress was made in studying large CF systems for the low-energy
excitations of the $\nu=1/(2p+1)$ states. Here the unprojected CF 
wave functions have the special feature that they have at 
most one electron in the second LL, and the projection operator
can be written simply as $K-\frac{1}{2}\hbar \omega_c$, where 
$K$ is the kinetic energy operator. This 
was exploited by Bonesteel \cite {Bonesteel}
to study large CF systems and to compute the gap to charged excitation.
The CF prediction for the gap was slightly lower than the 
one obtained from the Laughlin's trial wave functions.
The full dispersion of the  neutral excitation was determined by 
Kamilla, Wu and Jain using similar techniques \cite{Kamilla}.  
This approach, however, is not applicable to general fractions, for which
the number of electrons in higher LLs in the unprojected CF wave function
is a finite {\em fraction} of the total number of electrons, growing 
with the total number of electrons.

It is clearly desirable
to find ways of computing with the CF wave functions for large systems
at {\em arbitrary} filling factors. The study of $n/(2pn+1)$ states
at large $n$ is interesting not only in its own right but may also
shed further light on the compressible 1/(2p) state. 
In this paper, we report on a method which enables computations for
rather large CF systems. We give below results 
for as many as 50 composite fermions; treatment of 
bigger systems should be possible in the
future. This constitutes a significant step in our ability to achieve  
detailed quantitative predictions for the FQHE. 
Short reports on parts of this paper have been published elsewhere
\cite {JK,KJ}.

As a first application  of this method, we have 
computed the ground state energies
and the gaps to charged and neutral excitation for several FQHE 
states of interest.
There are several parameters in actual experiments, namely 
the finite quantum well width \cite {Wellwidth}, Landau level mixing
\cite {LLmixing}, finite Zeeman
energy, and disorder, which influence these energies.  
Our aim in this work will be to obtain a quantitative description 
in an idealized situation where these 
effects become unimportant, namely the limit in which the
transverse width of the quantum well, disorder, and $(e^2/\epsilon l)
/ (\hbar \omega_c)$ all vanish. The last has the effect of 
supressing any LL mixing or spin reversal. 
Besides being theoretically convenient and clean, this limit is 
sufficient for obtaining zeroth order estimates as well as trends. 
Furthermore, previous work tells us how to include these effects and how
much of a quantitative correction they make. 
For example, the gap to charged excitations
is reduced approximately by a factor of 2 when corrections due
to finite well width and LL mixing are taken into account for
typical sample parameters \cite {Zhang}. In fact, all of the effects neglected 
in the ideal limit are found to {\em reduce} the gaps, so 
our estimates will provide an upper bound for them.

Another promising approach toward a quantitative description of
the FQHE is the Chern-Simons field theoretic approach \cite {Lopez,HLR}.
It is best suited for a description of the long distance 
physics, an understanding of which is most crucial 
in the case of the Fermi sea of the composite fermions, 
where the correlations are expected to have  
power law decays. A computation
of the short distance properties like the gaps has not yet been 
possible in this approach, except asymptotically close to $\nu=1/2$ 
\cite {Stern}. 
In the most widely used framework \cite {HLR}, the CS field theory is 
treated as an {\em effective} long-distance theory of composite fermions, 
with the effective mass of composite fermions treated as an 
input parameter, to be fixed either by comparison with 
experiment or by other microscopic means.

In addition to providing a new calculational tool, the   
new lowest LL (LLL) representation of the CF wave functions also 
suggests how the CF theory can be motivated quite plausibly
within the
lowest Landau level, without appealing to higher LLs at all.
The composite fermions, their effective
magnetic field, their quasi-LL structure, and their effective
cyclotron energy are all seen to originate in this approach 
as a direct consequence of repulsive interelectron
interaction. The single particle wave functions of composite
fermions in the lowest (electronic) LL are shown to 
possess the appealing property of having zeros tightly
bound to electrons, aside from a few localized defects.
This shows that while the analogy to higher LLs is a 
convenient, and intuitively powerful route to the CF theory,
it is by no means necessary.

The organization of the rest of this chapter is as follows.
Section II gives the explicit form of the CF wave functions in the
lowest Landau level. The essential result is displayed in the beginning of
II.C in a form ready to be used by the interested reader. 
Section III presents results of our Monte Carlo calculations
for the ground state energies and the gaps.
In particular, it is found that the gaps to charged excitations are 
in reasonable agreement with the formula suggested 
by Halperin {\em et al.} \cite {HLR}, and the study of the 
neutral excitations uncovers an instability of FQHE for 
$\nu\leq 1/9$. In Section IV, we show how the composite
fermion wave functions can be motivated within the lowest LL, and
how concepts like effective magnetic field, effective filling
factor, CF-LL index, and effective cyclotron energy appear here.   
This work is formulated in the disk geometry, which also gives
us an excuse to compare the CF wave functions with the exact wave 
functions in the disk geometry. The paper is concluded in Section V. 

\section{Composite Fermion Wave functions in lowest Landau level} 
\label{cfwf}

We start by considering the spherical geometry 
in which $N$  electrons move on the two-dimensional
surface of a sphere \cite{Wu,Haldane} under the influence
of a radial magnetic field $B$. By virtue of having no 
boundaries, this geometry is convenient 
for an investigation of the bulk properties of the CF state.  
The disk geometry will be considered in a later section.

\subsection{Single electron states}

The single particle eigenstates are the monopole harmonics, $Y_{q,n,m}$
\cite{Wu}:
\begin{equation}
Y_{q,n,m}(\Omega)=N_{qnm} 2^{-m} (1-x)^{\frac{-q+m}{2}}
(1+x)^{\frac{q+m}{2}}  P^{-q+m,q+m}_{q+n-m}(x) 
e^{i(q-m)\phi}\;,  
\end{equation}
where $x=\cos \theta$, the 
total flux through the sphere is  equal to
$2q\phi_0$ ($2q$ is an integer),
$n=0,1,2,...$ is the LL index, $m=-q-n, -q-n+1, ..., q+n$
labels the degenerate states in the $n$th LL, and $l=q+n$.
$\theta$ and $\phi$ are the usual angular coordinates of
the spherical geometry, and $P^{\alpha,\beta}_{\gamma}$ is the
Jacobi polynomial.  The normalization coefficient is given by
\begin{equation}
N_{qnm}=\left(\frac{(2q+2n+1)}{4\pi}\frac{(q+n-m)!(q+n+m)!}{n!(2q+n)!}
\right) ^{1/2}\;\;.
\end{equation} 
Substituting the explicit form for the Jacobi polynomial, 
$Y$ can be rewritten as\footnote{Throughout this work, the 
binomial coefficient
${{\gamma \choose \beta}}$ is to be set equal to
zero if either $\beta >\gamma$ or $\beta<0$.} 
\begin{equation}
Y_{q,n,m}(\Omega)=N_{qnm} (-1)^{q+n-m} e^{iq\phi_j} u_j^{q+m} v_j^{q-m}
\sum_{s=0}^{n}(-1)^s {{n \choose s}} {{ 2q+n \choose q+n-m-s}}
 (v^*v)^{n-s}(u^*u)^s\;\;,
\label{mh}
\end{equation}
where $\Omega$ represents the angular coordinates 
$\theta$ and $\phi$ of the electron, and 
\begin{equation}
u\equiv \cos(\theta/2)\exp(-i\phi/2)
\end{equation}
\begin{equation}
v\equiv \sin(\theta/2)\exp(i\phi/2)\;.
\end{equation} 

\subsection{Composite fermion states -- unprojected}

Now we consider interacting electrons at flux $Q$.  
The composite fermion theory \cite {Jain} postulates that the 
strongly correlated 
liquid of interacting electrons at $Q$ 
is equivalent to a weakly interacting
gas of composite fermions at $q=Q-p(N-1)$. 
(We omit the conventional superscript $CF$ or $*$ on $q$ to simplify 
the formulas.)
The unprojected wave function \cite {Kamilla} for the CF state at 
$q$, $\Phi^{CF-up}$,
is related to that of non-interacting  
electrons at $q$, $\Phi$, as 
\begin{equation}
\Phi^{CF-up}={\cal J}\;\Phi\;\;;
\end{equation}
the Jastrow factor ${\cal J}$ is given by 
\begin{equation}
{\cal J}\equiv \Phi_{1}^{2p}=\prod_{j<k}
(u_j v_k-v_j u_k)^{2p} \exp[ip(\phi_{j}+\phi_k)]\;,
\end{equation} 
where $\Phi_1$ is the wave function of the lowest filled LL.

\subsection{Lowest LL projection}

Since we are interested in the limit of $B\rightarrow \infty$,
we will be using the LLL projections of $\Phi^{CF-up}$,
denoted by $\Phi^{CF}$:
\begin{equation}
\Phi^{CF} \equiv {\cal P} \Phi^{CF-up}\;,
\end{equation}
where ${\cal P}$ is the LLL projection operator.
We begin by quoting the final result \cite {JK} in a form that can
be used directly by someone not interested in the details of
its derivation, which fill the rest of the subsection. 

{\bf Final Result:}

$\Phi$, the wave function of
electrons at $q$, is a linear superposition of Slater determinant
basis states made of $Y_{q,n,m}$.
The corresponding wave function of composite
fermions at $q$ (which gives the wave function of interacting
electrons at $Q$) is same as $\Phi$, except with $Y_{q,n,m}$ replaced
by $Y^{CF}_{q,n,m}$, given by 
\begin{eqnarray}
&&Y_{q,n,m}^{CF}(\Omega_j) =
N_{qnm} (-1)^{q+n-m} \frac{(2Q+1)!}{(2Q+n+1)!}
u_j^{q+m}   v_j^{q-m}  \nonumber \\
&&
\sum_{s=0}^{n}(-1)^s {{n \choose s}} {{ 2q+n \choose q+n-m-s}}
\; u_j^s \;    v_j^{n-s}  \; \left[ \left({\frac{\partial}{\partial u_j}}
\right)^s 
\; \left({\frac{\partial}{\partial v_j}}\right)^{n-s} J_j^p\right]\;
\end{eqnarray}
where
\begin{equation}
J_j=\prod_{k}^{'} 
(u_j v_k-v_j u_k)\exp[\frac{i}{2}(\phi_{j}+\phi_k)].
\end{equation}
Here the prime denotes the condition $k\neq j$.

This implies that $Y_{q,n,m}^{CF}(\Omega_j)$ can be interpreted  
as the single particle basis functions of composite fermions, 
using which many-CF states can be constructed in the usual 
manner, in complete analogy to the non-interacting electron 
states.  

In order to deal with the derivatives in our Monte 
Carlo calculations, we find it convenient to 
write them as follows:
\begin{equation}
\left({\frac{\partial}{\partial u_j}}\right)^s
\; \left({\frac{\partial}{\partial v_j}}\right)^{n-s} 
J_j^p= J_j^p \left[ \overline{U}_j^s \overline{V}_j^{n-s} 1 \right]
\end{equation}
where 
\begin{equation}
\overline{U}_j=J_j^{-p} \frac{\partial}{\partial u_j} 
J_j^{p}=p \sum_{k}^{'}\frac{ v_k}{
u_j v_k - v_j u_k} + \frac{\partial}{\partial u_j}\;\;,
\end{equation}
\begin{equation}
\overline{V}_j= J_j^{-p} \frac{\partial}{\partial v_j} 
J_j^{p}=p\sum_{k}^{'}\frac{
-  u_k}{u_j v_k - v_j u_k} + \frac{\partial}{\partial v_j}\;.
\end{equation}
For a given $n$, explicit analytical form of the derivatives
is used in the evaluation of the wave function.
The advantage of this form is that the $J_j$'s can be factored
out of the Slater determinants to give back the usual Jastrow 
factor 
\begin{equation}
{\cal J}=\prod_j J_j^{p}\;.
\end{equation}
Note that the phase factors $\exp[\frac{i}{2}(\phi_{j}+\phi_k)]$
do not play any role in the calculation of the energies.
 
{\bf Derivation:}

To derive this result, we consider a case 
when $\Phi$ is a single Slater determinant,
given by 
\begin{equation}
\Phi=Det[Y_i(\Omega_j)]=\left | \begin{array}{ccccc}
                  Y_{1}(\Omega_1) & Y_{1}(\Omega_2) & . & . & . \\
                  Y_{2}(\Omega_1) & Y_{2}(\Omega_2) & . & . & . \\
                  . & . & . &  &  \\
                  . & . &  & . &  \\
                  . & . &  &  & . 
                \end{array} \right |
\label{phi}
\end{equation}
so the unprojected CF wave function is given by
\begin{equation}
\Phi^{CF-up}= {\cal J} Det[Y_i(\Omega_j)]\;.
\end{equation}
A generalization to situations in which $\Phi$ is a linear 
superposition of many Slater determinants is straightforward.
We need to show that $\Phi^{CF}$ is obtained by replacing 
$Y$'s in Eq.~(\ref{phi}) by $Y^{CF}$'s.

The Jastrow factor can be incorporated into the Slater determinant
to give  
\begin{equation}
\Phi^{CF-up}=Det[Y_{i}(\Omega_j)J_j^p] \;.
\end{equation}
We now project this on to the lowest LL by projecting each 
element onto the lowest LL. I.e., we write
\begin{equation}
\Phi^{CF}=Det[{\cal P}Y_{i}(\Omega_j)J_j^p] \;.
\label{phicfs}
\end{equation}
Since the product of two LLL wave functions is also in the lowest LL, 
$\Phi^{CF}$ is guaranteed to be in the lowest LL.
It remains now to show that 
\begin{equation}
{\cal P}Y_{q,n,m}(\Omega_j)J_j^p=Y^{CF}_{q,n,m}(\Omega_j)\;.
\label{eq29}
\end{equation}

In order to evaluate the projection we first show that
there exists an operator
$\overline{Y}^{q'}_{q,n,m}$ satisfying the property that 
\begin{equation}
{\cal P} Y_{q,n,m} Y_{q',0,m'}= \overline{Y}^{q'}_{q,n,m} Y_{q',0,m'}\;\;, 
\label{op}
\end{equation}
where $Y_{q',0,m'}\sim e^{iq'\phi_j}u_j^{q'+m'}
v_j^{q'-m'}$ is a LLL wave function for monopole 
strength $q'$. For the present purposes, it 
is important that $\overline{Y}^{q'}_{q,n,m}$ be independent
of $m'$ although it may depend on $q'$; this is because
the single particle wave functions of the $j$th electron appearing
in $J_j$ have different $m'$ but the same $q'=(N-1)/2$.  
To this end, we multiply one of the terms on the right hand
side of Eq.~(\ref{mh})
by the LLL wave function $Y_{q',0,m'}$
and write (with $Q\equiv q+q'$, $M\equiv m+m'$, 
and the subscript $j$ suppressed):
\begin{equation}
e^{iQ\phi} (v^*v)^{n-s}(u^*u)^s u^{Q+M}v^{Q-M}=a_0
e^{iQ\phi}u^{Q+M}v^{Q-M}+ higher\;\; LL \;\; states.
\label{proj}
\end{equation} 
For $|M|>Q$, $a_0$ must vanish, since $|M|\leq Q$ in the lowest LL.
Let us first consider the case $|M|\leq Q$. Multipling both sides
by $e^{-iQ\phi} u^{*Q+M}v^{*Q-M}$ and integrating over the angular 
coordinates gives
\begin{equation}
a_0=\frac{(Q+M+n-s)!(Q-M+s)! (2Q+1)!}{(Q+M)! (Q-M)! (2Q+n+1)!}\;.
\end{equation}
This shows that, apart from an $m'$-independent 
multiplicative constant $(2Q+1)!/(2Q+n+1)!$, the LLL projection of 
the left hand side of Eq.~(\ref{proj}) can 
be accomplished by first bringing all $u^*$ and $v^*$ to the left and
then making the replacement
\begin{equation}
u^*\rightarrow \frac{\partial}{\partial u}\;,
\;v^*\rightarrow \frac{\partial}{\partial v}\;.
\end{equation}
While true in general for $|M|\leq Q$, 
this prescription can be shown to produce the correct result  
(i.e., zero)
even for $|M|>Q$ for the LLL projection of states of the form
$Y_{q,n,m} Y_{q',0,m'} $, as proven in the Appendix.
Thus, in Eq.~(\ref{op}), 
\begin{eqnarray}
\overline{Y}^{q'}_{q,n,m}
&=& \frac{(2Q+1)!}{(2Q+n+1)!}
N_{qnm} (-1)^{q+n-m} e^{iq\phi_j} \nonumber \\
&&\sum_{s=0}^{n}(-1)^s {{n \choose s}} {{ 2q+n \choose q+n-m-s}}
\left(\frac{\partial}{\partial u}\right)^{s} u^{q+m+s} 
\left(\frac{\partial}{\partial v}\right)^{n-s} v^{q-m+n-s}\;.
\label{fleft}
\end{eqnarray}

A delightful simplification occurs when one brings all the 
derivatives to the right in Eq.~(\ref{fleft}) using  
\begin{equation}
\left(\frac{\partial}{\partial v}\right)^{\beta} v^\gamma =
\sum_{\alpha=0}^{\beta} \frac{\beta!}{\alpha !} {{\gamma \choose \beta-
\alpha}} v^{\gamma-\beta+\alpha}
\left(\frac{\partial}{\partial v}\right)^{\alpha}\;,
\end{equation}
and a similar equation for the derivative with respect to $u$ (with 
the summation index $\alpha'$).
The sum over $s$ in Eq.~(\ref{fleft}) then takes the form
\begin{equation}
\sum_{s=\alpha'}^{n-\alpha} (-1)^{s} {{n-\alpha-\alpha' \choose 
s-\alpha'}}=\sum_{s'=0}^{n-\alpha-\alpha'}(-1)^{\alpha'+s'}
{{n-\alpha-\alpha' \choose s'}} \;,
\end{equation}
which is equal to $(-1)^{\alpha'}(1-1)^{n-\alpha-\alpha'}$ and 
vanishes unless $n=\alpha+\alpha'$.  The only term 
satisfying this condition is one with $\alpha=n-s$
and $\alpha'=s$. Consequently, the derivatives in Eq.~(\ref{fleft}) 
can be moved to the extreme right and act only on the following 
LLL wave function. This gives Eq.~(\ref{eq29}), completing our proof. 

Before ending this subsection, we note that with the help of the
identity $u^*u+v^*v=1$, the Eq.~(\ref{mh}) can be cast into
different equivalent forms; for example, the sum over $s$ can 
be expressed as a power series entirely in $v^*v$ or $u^*u$. 
While the projection can be carried out with 
any form, the subsequent formulas are simplest expression  
in Eq.~(\ref{mh}) is used.
Otherwise, the simple replacement of $u^*$ or $v^*$ by the 
corresponding derivatives does not give the projection
and neither are the cancellations indicated above obtained. 

\section{Low energy states}

The most important goal of theory is to provide an understanding
of the low-lying states of the system in question. We will
concentrate here on the ground and the low-energy states at
the special filling factors $\nu=n/(2pn+1)$. 
The ground state energy of the incompressible states is
of interest, even though it cannot be measured directly, 
because it serves as  
an important criterion when comparing different theories of
the FQHE, or in investigating the question of the stability of
a FQHE state relative to another state with a different symmetry, 
e.g., the Wigner crystal \cite {Wigner,CWC}. Two kinds of 
excitations are relevant for experiments: the charged and the 
neutral excitations. The gap to the charged excitations appears 
in transport experiments as the activation energy, while
the dispersion of the neutral excitations, in particular 
the gap at the minima or maxima, can be measured by inelastic  
light \cite {Pinczuk} or phonon scattering \cite {Mellor}.
 The smallest energy required to 
create an excitation  will also play an important role in the 
low-temperature thermodynamics of the FQHE. 

Let us first summarize the status of our quantitative 
understanding of the FQHE prior to the the CF approach. 
(i) Ground state energy: The energies of the 1/3 and 2/5 states
have been determined reasonably accurately from an  
extrapolation of exact diagonalization results \cite {Fano,AMorf}.
A good {\em ansatz} exists for the ground state
at $\nu=1/(2p+1)$ in the form of the Laughlin \cite {Laughlin}  
wave functions. The energy of the Laughlin state
is known accurately from Monte Carlo study \cite {Morf} of large systems,
and is in good agreement with that obtained from the exact 
diagonalization results. 
(ii) Charge gap: Almost the same story holds here. Reasonable 
estimates are available for the gaps of 1/3, 2/5 and, to some
extent, for 3/7 from exact diagonalization studies \cite {AMorf}.
The gap of the 1/3 state has also been calculated using 
the Laughlin trial wave functions.
(iii) Neutral excitations: A single mode approximation 
(SMA) \cite {GMP} was found to work reasonably
well for 1/3, but less well for other Laughlin states, as we shall 
see, and not well for other FQHE states. 
The exact diagonalization studies exhibit strong finite size 
effects for the $n/(2n+1)$ states with 
$n\neq 1$ and fail to give a reliable 
account for the dispersion of the neutral excitations. 

In the CF theory, the FQHE ground state at $\nu=n/(2pn+1)$
corresponds to $n$ filled CF-LLs. The excitations 
are obtained by removing  
one composite fermion from the $n$th CF-LL, leaving behind
what we will call a ``quasihole", and placing it in the 
$(n+1)$st CF-LL, creating a ``quasiparticle". This excitation
will be called a (CF-)exciton. The orbital 
angular momentum $L$ of the state is related  
to the wave vector $k$ of the planar geometry as
$L=k R$ where $R=\sqrt{Q}$ is the radius of the sphere,
in units of the magnetic length ($l\equiv \sqrt{\hbar c/eB}$).
The configuration with the quasiparticle and quasihole  
on opposite poles will be identified with the 
charged excitation, as the distance between the quasiparticle and
the quasihole is the maximum; this corresponds to 
the largest $L$ state of the exciton.

The transport gaps at $\nu=1/(2p+1)$ have been computed using 
the CF theory \cite {Bonesteel,others}.  It was found that
the CF theory gives marginally smaller gaps than those computed 
using Laughlin's trial wave functions (the difference
is more significant for short range interactions than for 
the long range Coulomb interaction). We note here that here
the CF wave functions for the ground state and the quasihole are
identical to those of Laughlin's, but the quasiparticle wave
functions are different, resulting in unequal values of the gaps. 

\subsection{Calculation Procedure}

The interaction energy per particle is
\begin{equation}
V=\frac{1}{N}
\sum_{i<j} \frac{e^2}{\epsilon R_{ij}}\;,
\end{equation}
\begin{eqnarray}
R_{ij} \; = &\; & 2R|\cos(\theta_i/2)\sin(\theta_j/2) \; -  \;
\cos(\theta_j/2)\sin(\theta_i/2)e^{i(\phi_i-\phi_j)}|\;,
\end{eqnarray}
where $R_{ij}$ is the arc distance between two electrons,
and $R=\sqrt{Q} \;l$ is the radius of the sphere.
It will be evaluated by variational
Monte Carlo techniques, with the absolute value
squared of the Slater determinantal wave function
of composite fermions used as the probability weight function for the
Metropolis acceptance sampling.  The methods are standard, but 
a few points should be mentioned.

It takes ${\cal O}(N^3)$ operations to evaluate a Slater  
determinant. For a free Fermion problem, if a single particle
is moved, only one row (or column) of the density matrix is altered,
and it takes only $N$ operations to evaluate the new
determinant in terms of the old inverse matrix and
${\cal O}(N^2)$ operations to update the inverse matrix if the move is
accepted \cite {Ceperley}. However, the CF Slater determinant
describes a {\it strongly correlated } electron system and
hence is explicitly {\it non-local}, i.e. when a single electron is
moved all CF matrix elements are altered, since the
single particle wave function of a given composite fermion
depends on the positions of all the particles.
Therefore, an update of the trial $N \times N$ density matrix of the
CF system proceeds through ${\cal O}(N^3)$ operations at every
Metropolis step.                           

The time $T$ involved in our variational Monte Carlo
calculations for the coulomb interaction energies of the $N$ -
electron system in the $n \over |2pn \pm 1|$
FQH state scales as:
$$T \;\; \propto \;\; {N^3}{\epsilon}^{-2}n^3 $$
the cost of evaluating the
CF determinant goes as ${N^3}$,  the cost of
projection as $n^3 $, and  
with $\epsilon = {\displaystyle
{1 \over \sqrt{N_{mc}}}}$, where $\epsilon $ is the desired
simulation error and $N_{mc}$ is the total number of Monte
Carlo steps. (The Monte Carlo statistical error scales as
$\sigma_{\psi}/\sqrt{N_{mc}/\xi_c}$, where $\sigma_{\psi}$ is the
standard deviation in the wave function 
and $\xi_{c}$ is the correlation time, which is
a property of the wave function being sampled.)
Typically, we have performed calculations on systems
containing upto 50 particles. For example, a calculation of the
ground state interaction energy on the $\nu=2/5$ CF state for
30 particles with about 20 million Monte Carlo steps
takes $\sim 30$ CPU hours on our work-station
(DEC Model 250/4, CPU 21064A, Mhz 266).
A similar calculation for 30 particles in the $\nu=3/7$ CF
state takes $\sim 100$ CPU hours on the same machine.

For the ground states and the charged excitations at $\nu=n/(2pn+1)$
the CF wave functions are single
Slater determinants. For the neutral exciton, the wave function 
is a linear superposition of $\sim N/n$ determinants, making the   
computations significantly more time consuming.  In addition, 
the energy of the exciton has to be evaluated with extremely
high accuracy
in order to obtain reasonable values for the energy differences;
up to $4\times 10^8$ Monte Carlo steps were carried out in our
computations of the exciton energy. A good accuracy is especially
important at small $\nu$, where the exciton energy 
becomes rather small;
we will be dealing with energy differences as small as
0.001 $e^2/\epsilon l$, which is approximately two orders of
magnitude smaller than the transport gap of the 1/3 state.

We note that when Coulomb 
energies are calculated for finite systems with an intent to obtain
thermodynamic estimates, usually an infinite Ewald summation 
\cite {Ceperley} has to be
performed that accounts for the interaction of the charges with
their image charges with respect to the finite boundary imposed in
the calculation; no such sum is necessary in the spherical geometry
due to absence of boundaries.

\subsection{Accuracy of CF wave functions}

Before we proceed to compute the 
energies, we ask how reliable the CF wave functions are.   
To this end, we compare the energies of the CF wave functions to 
the exact eigenenergies obtained numerically, both for the
ground state and the charged excitation, shown in 
Table I. The exact energies were obtained 
by a numerical diagonalization of the Hamiltonian in the lowest 
LL subspace, assuming an ideal two-dimensional system.
Further details can be found in the literature (see, for example,
Wu and Jain \cite {Dev}). All energies quoted in Table I and 
the rest of the paper include the interaction with the 
positively charged neutralizing background, which amounts to 
an addition of $-N^2/2\sqrt{Q}$, the interaction 
energy of the $N$ electrons with a positive charge   
$+Ne$ at the center of the sphere.
The accuracy of the CF theory is impressive; the 
predicted energies agree with the exact energies to better 
than  $0.05\%$ for the systems studied, which represent 
the biggest systems for which exact diagonalization has been  
performed at 2/5 and 3/7. We note that these system sizes are 
already large 
compared to the relevant length in the problem, namely the
magnetic length (or the effective magnetic length), 
which, coupled with the fact that the energies are determined largely
by the short-distance behavior, suggests that the error in the 
thermodynamic limit ought not to be much larger. 
In other words, extrapolation of the CF energies to large $N$
will produce virtually the same results as extrapolation of the 
exact energies. 

Since the gap is an $O(1)$ energy obtained as the
difference between two large $O(N)$ energies, its accuracy 
is expected to be less than that of either the ground 
state or the excited state energy. 
A comparison of the gaps predicted by the CF theory with 
with the corresponding exact gaps (Table I) shows that the 
discrepancy is of the order of a few percent.  This level of 
accuracy is possible only because the ground and the 
excited state energies are extremely accurate.
The full dispersions of the neutral excitations obtained in the
CF theory have a similar level of accuracy \cite {Kamilla}. 

It is obvious that the ground state wave functions considered
here contain {\em no adjustable parameters} as they are related
to the unique filled LL wave functions. Though less obvious, this
is also true for the single exciton state as well (at 
arbitrary wave vector); they are also completely
determined by symmetry in the restricted Hilbert
space of the CF wave functions.  
Also, since the CF wave functions constructed here are strictly
within the lowest {\em electronic} LL, their energies 
provide strict variational bounds in the limit 
of zero thickness, no disorder, and large magnetic
field. 

Another point that needs to be emphasized is that although 
the CF wave functions are analogous to, and indeed
motivated by  the wave functions of {\em non-interacting} electrons
at $q$, they describe {\em fully interacting} electrons at $Q$.
In other words, {\em the CF wave functions provide an account of the 
residual interaction between composite fermions themselves}.
In this sense, the wave functions go beyond the simple mean-field 
picture of non-interacting composite fermions. We will 
see below that the residual interaction between composite fermions 
is responsible for the destabilization of the FQHE 
for $\nu\leq 1/9$ \cite {KJ}. 

\subsection{FQHE Ground states}

\begin{figure}
\centerline{\psfig{file=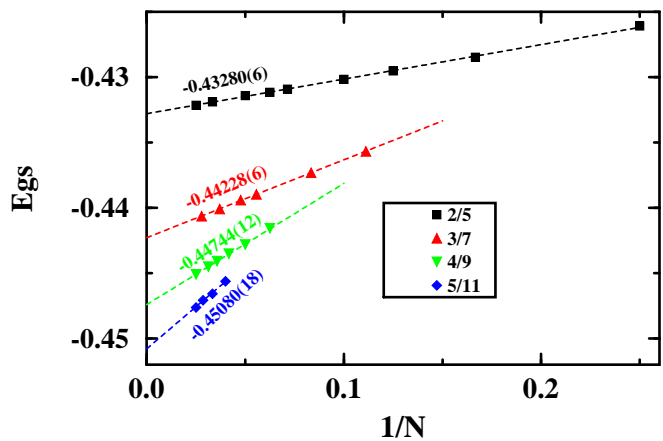,width=4in,angle=-90}}
\centerline{\psfig{file=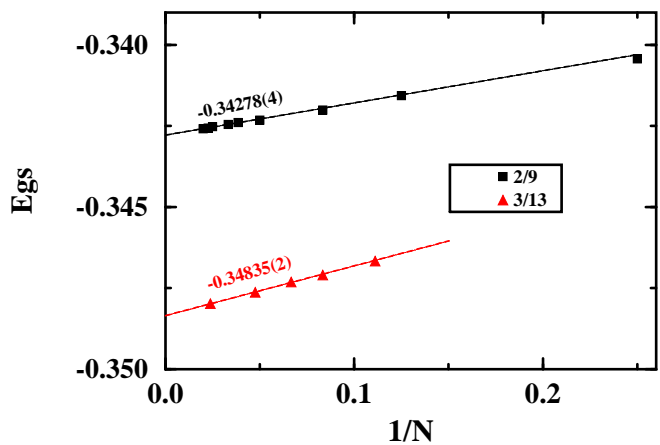,width=4in,angle=-90}}
\caption{Ground state energy of various FQHE states as a function of
the number of particles, $N$.}
\label{fig:rev_Fig1}
\end{figure}

Having estimated the level of confidence in the CF theory, we
proceed to study larger systems. 
The energies of the CF ground states for several values of $N$
at 2/5, 3/7, 4/9, 5/11, 2/9, 3/13 and 4/17 are given in Table III.
While estimating the $N\rightarrow \infty$ limit, we
find it useful to correct these energies for the finite size
deviation of the electron density from its thermodynamic value,
$\rho$, by multiplying the energies by a factor
$\sqrt{\rho/\rho_{N}}=\sqrt{2Q\nu/N}$, following Ref.~\cite {Morf}.
This substantially reduces $N$ dependence 
and thereby allows for better extrapolation to the thermodynamic
limit, shown in Fig. \ref {fig:rev_Fig1}.  The thermodynamic estimates are  
obtained using a chi-square fitting of the energy as a 
polynomial in $1/N$ that biases the points by
their error bars; the number quoted in Table III is an average 
over estimates obtained through linear, quadratic, and cubic 
extrapolation.

\begin{figure}
\centerline{\psfig{file=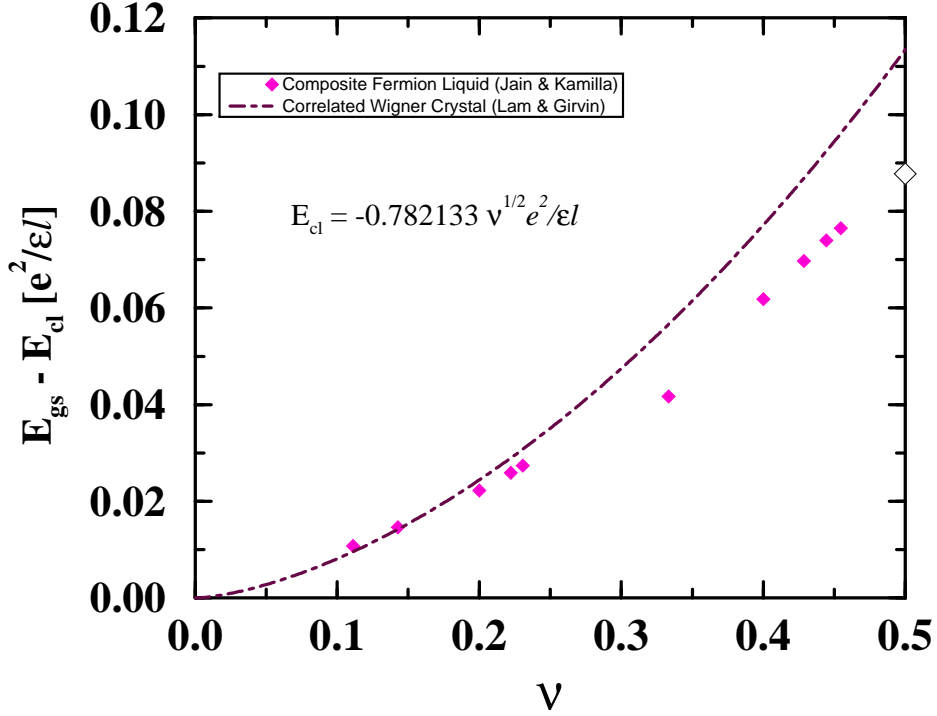,width=6in,angle=-90}}
\caption{The energy of the liquid FQHE states and the correlated WC state.
The WC energy is taken from \protect \cite {Lam}. Only the quantum
correction to the energy of the classical ground state
(a two-dimensional Wigner crystal with triangular symmetry) is shown, where
{\protect $E_{cl} = -0.782133 \protect \sqrt{\nu}$ }
\protect $e^2/\epsilon l$.  The open diamond on the right
vertical axis is our estimate of the CF ground state energy at
$\nu = 1/2$, $(E_{gs,1/2} - E_{cl} = 0.0877(2) e^2/\epsilon l)$
obtained by extrapolating the energies of the $n/(2n+1)$ CF states.
Also note that the energy of the CF liquid is shown only at the
special \protect $n/(2pn+1)$ filling factors; the full curve will have
cusps at these points. }
\label{fig:rev_Fig2}
\end{figure}

Fig. \ref{fig:rev_Fig2} shows the ground state energy of the CF liquid state along
with the best estimate for the energy of 
the Wigner crystal \cite {Lam}. It  illustrates nicely the 
fundamental reason for why the 
Wigner crystal state is not observed as soon as the kinetic
energy is suppressed by forcing all electrons to occupy the lowest 
LL: it is preempted by the CF liquid
which has substantially lower energy. We note here that while
the energies plotted in Fig. \ref {fig:rev_Fig2} are variational estimates, they are believed 
to be extremely accurate approximations to the actual energies, so,
the conclusion drawn from this figure is reliable  
except in the crossover region near $\nu=1/7$, where
the two energies are very close. We will have more to say in this
regard later; here we would like to stress that our estimates  
seem to rule out Wigner crystal in the vicinity of $\nu=1/2$.

Due to particle-hole symmetry in the lowest LL,
the ground state energy per particle of the state at 
\begin{equation}
\nu'=1-\frac{n}{2pn+1}\;\;
\end{equation}
is related to the ground state energy per particle of the state at $
\nu=\frac{n}{2pn+1}$ as  
\begin{equation}
\nu [E_g(\nu) - E_g(1)] = (1-\nu)[E_g(1-\nu) + E_g(1)],
\end{equation}
where $E_g(1) = -\sqrt{\pi/8}$, is the Hartree-Fock energy
of a filled Landau level, and all energies are in units of 
$e^2/\epsilon l$ \cite {Fano2}.

\subsection{Transport Gaps}

\begin{figure}
\centerline{\psfig{file=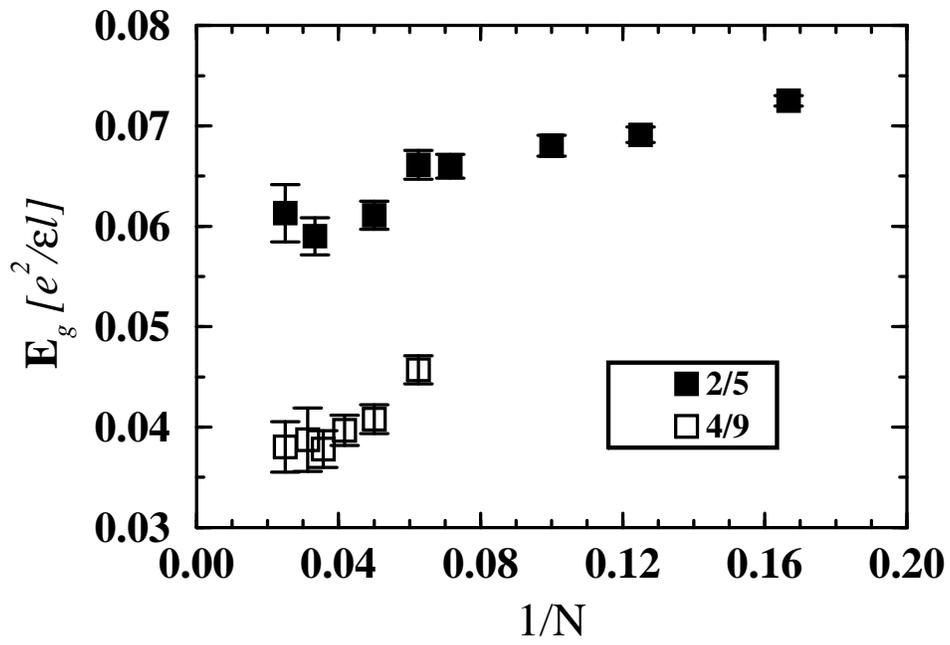,width=6in,angle=-90}}
\caption{$N$ dependence of the ``corrected'' gap for 2/5 and 4/9.}
\label{fig:rev_Fig3}
\end{figure}

Next we come to the energy of the charged 
excitation (the CF exciton with the largest $L$). 
For $\nu=1/(2p+1)$, the gap computed with the present projection
technique is identical to that computed with the full projection,
investigated earlier by Bonesteel \cite {Bonesteel}, and we
will simply use his results. The gaps for several values of $N$
at 2/5, 3/7, 4/9, 5/11, 2/9, 3/13 and 4/17 are given in Table III.
We subtract from the gap $-(2p+1)^{-2}/(2\sqrt{Q})$, 
the interaction between pointlike particles of  
charges $e/(2p+1)$ and $-e/(2p+1)$ placed at the two poles,
to correct for the interaction between the quasiparticles; these
are indicated as the corrected gap in Table III (no correction
is made for the density). Fig. \ref{fig:rev_Fig3} shows 
typical dependence of the gap on the number of particles $N$. 
The $N\rightarrow \infty$ estimates of the gaps are given in 
Table III. We note that the gaps of two states related by 
particle-hole symmetry are the same, when measured in units of 
$e^2/\epsilon l$. 

Apart from the intrinsic error of the CF wave functions, 
believed to be quite small, there are two sources of uncertainty 
in our estimates, both of which can be further reduced,
if so needed. One is, of course, the statistical 
error in our Monte Carlo, and the other comes from the 
non-monotonous, oscillatory part of the interaction 
between the quasiparticle and the quasihole as a function 
of their distance, originating 
from the fact that the quasiparticles are not 
point-like but extended in space and their density 
exhibits oscillations as a function of the distance from the 
center, analogous to the LL wave functions.
As a result, the interaction energy becomes  
equal to the correction term used above 
only when the distance is very large compared to the size 
of the quasiparticles, which scales as $\sim \sqrt{2n}$ 
times the {\em effective} magnetic length, requiring a 
study of bigger and bigger systems as we go to larger
$n$ along the $n/(2n+1)$ sequence.
For 4/9 or 5/11, such fluctuations can be
seen even in systems with as many as 40 fermions.
A reduction of this error will require 
investigations of larger systems.

\begin{figure}
\centerline{\psfig{file=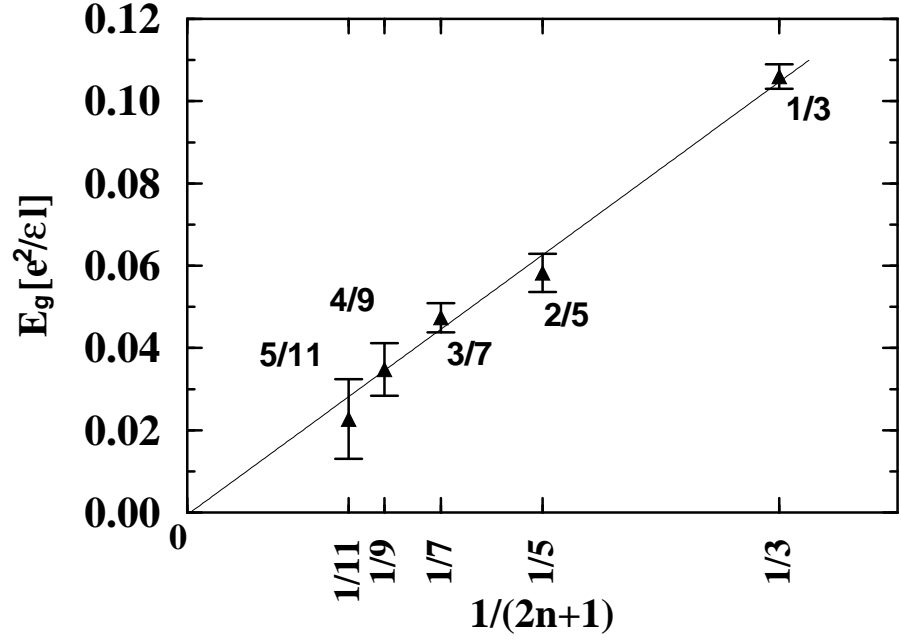,width=6in,angle=-90}}
\caption{Thermodynamic values of the gaps plotted as a function
of $1/(2n+1)$.
The results for the 1/3 gap are taken from Ref.\protect \cite {Bonesteel}.
The straight line is obtained by a chi-square fit, given by
\protect $E_g=0.000(5) + 0.316(21) (2n+1)^{-1}$.}
\label{fig:rev_Fig4}
\end{figure}

In a mean-field picture, it is natural to equate the transport gap to 
the effective cyclotron energy of composite fermions. This would give
\begin{equation}
E_{g}=\hbar \frac{e B^{CF}}{m^{CF} }=\hbar \frac{e B}{(2pn+1)m^{CF} }\;\;.
\end{equation}
However, since the gap in the LLL constrained theory must be 
proportional to $\sqrt{B}$, the effective mass of composite
fermions cannot be a constant but rather must be $m^{CF}\sim \sqrt{B}$. 
This kind of reasoning led Halperin {\em et al.} \cite {HLR} 
to conjecture that the gaps  of the $n/(2pn+1)$ states are given  
by 
\begin{equation}
E_{g}=\frac{C}{(2pn+1)}\frac{e^2}{\epsilon \ell}\;\;,
\label{gap}
\end{equation}
where it was estimated from the then existing results that 
the constant $C\approx 0.31$. 
As the Fig. \ref{fig:rev_Fig4} shows, the overall behavior of the gaps computed here 
is in good agreement with Eq.~(\ref{gap}) for the $p=1$ sequence.  
When the gaps are plotted in constant 
units (say, Kelvin) as a function of $(2n+1)^{-1}$, the best 
straight line fit has a negative intercept. A fit as a function
of $1/n$ is also not as satisfactory. Thus, our calculations
provide a microscopic justification for Eq.~(\ref{gap}).
For the $p=2$ FQHE states, the fit is less satisfactory. 
The analysis of Stern and Halperin \cite {Stern} predicts  
logarithmic corrections to the formula in Eq.~(\ref{gap}), 
which however are small except in the close 
vicinity of the half filled LL. At the level of the accuracy 
of our study, we do not expect to be able to 
distinguish such small deviations.

At the moment, it is not possible to compare the values of 
the gaps obtained here with experiments, due to the neglected
effects of finite
width of the quantum well Landau level
mixing and disorder. 
Nevertheless, the trend predicted by the CF theory, as summarized 
in Eq.~(\ref{gap}), is in 
good agreement with the experimental behavior.
The experimental data \cite {gaps} are reasonably well described by
\begin{equation}
E_{g}=\frac{C'}{(2n+1)}\frac{e^2}{\epsilon \ell}-\Gamma\;\;,
\end{equation}
which suggests that 
the finite width and LL mixing might, to zeroth order, only alter
the value of $C$, and disorder might enter in the form of
a constant shift of the gaps, which was interpreted by Du {\em et al.}
as the disorder induced broadening of the CF-LLs. 

Our work can be generalized to include some of the effects
neglected in the system considered above.
The finite well width softens the Coulomb interaction at short 
distances, which is straightforward to incorporate into our Monte
Carlo and will be the subject of a future work. 
Landau level mixing is harder to deal with.
It has been investigated by using Jastrow factors that
are not analytic in $z_j$ and thereby contain some LL mixing. 
Fixed phase Monte Carlo \cite {Ceperley2} or a variational approach 
considering a linear combination of $\Phi^{CF-up}$ and $\Phi^{CF}$ 
\cite {Bonesteel2} will be useful in this context.

\subsection{Neutral excitons}

Next we come to neutral excitations \cite {Kamilla}. 
Consider a state with a hole in 
the $m_h$ state of the topmost occupied
CF-LL  and a composite fermion in the $m_e$ state of the
lowest unoccupied  CF-LL. With no loss of generality, we assume
$\sum m_j=0$ for the excited state, i.e. $m_h=m_e$, and denote the
corresponding Slater determinant basis state by $|m_h>$.
Then, the exciton state with a definite total angular momenta $L$ is
a linear superposition of $|m_h>$ with different $m_h$, given by
\begin{equation}
\chi_{L}^{exciton}=\sum_{m_h}<n_t+q,-m_h;n_t+1+q,m_h|L,0>|m_h>\;,
\end{equation}
where $n_t$ is the LL index of the topmost occupied CF LL, and
$m_h= -q-n_t, ..., q+n_t$. Note that the coherence factors 
are the same as those for the IQHE exciton. 

\begin{figure}
\centerline{\psfig{file=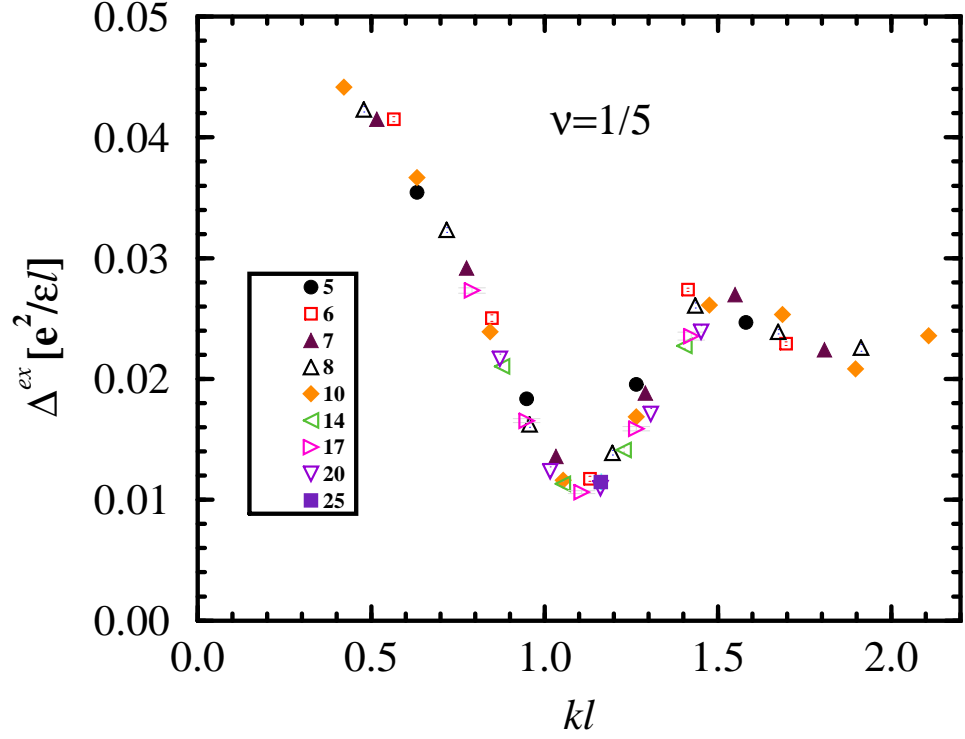,width=6in,angle=-90}}
\caption{Exciton dispersions for FQHE states at $\nu=$
1/5 for systems with several sizes, with $N$ shown
on the figures. All curves assume zero thickness.
The error bars are
shown explicitly whenever they exceed the sizes of the symbols. }
\label{fig:rev_Fig5}
\end{figure}

\begin{figure}
\centerline{\psfig{file=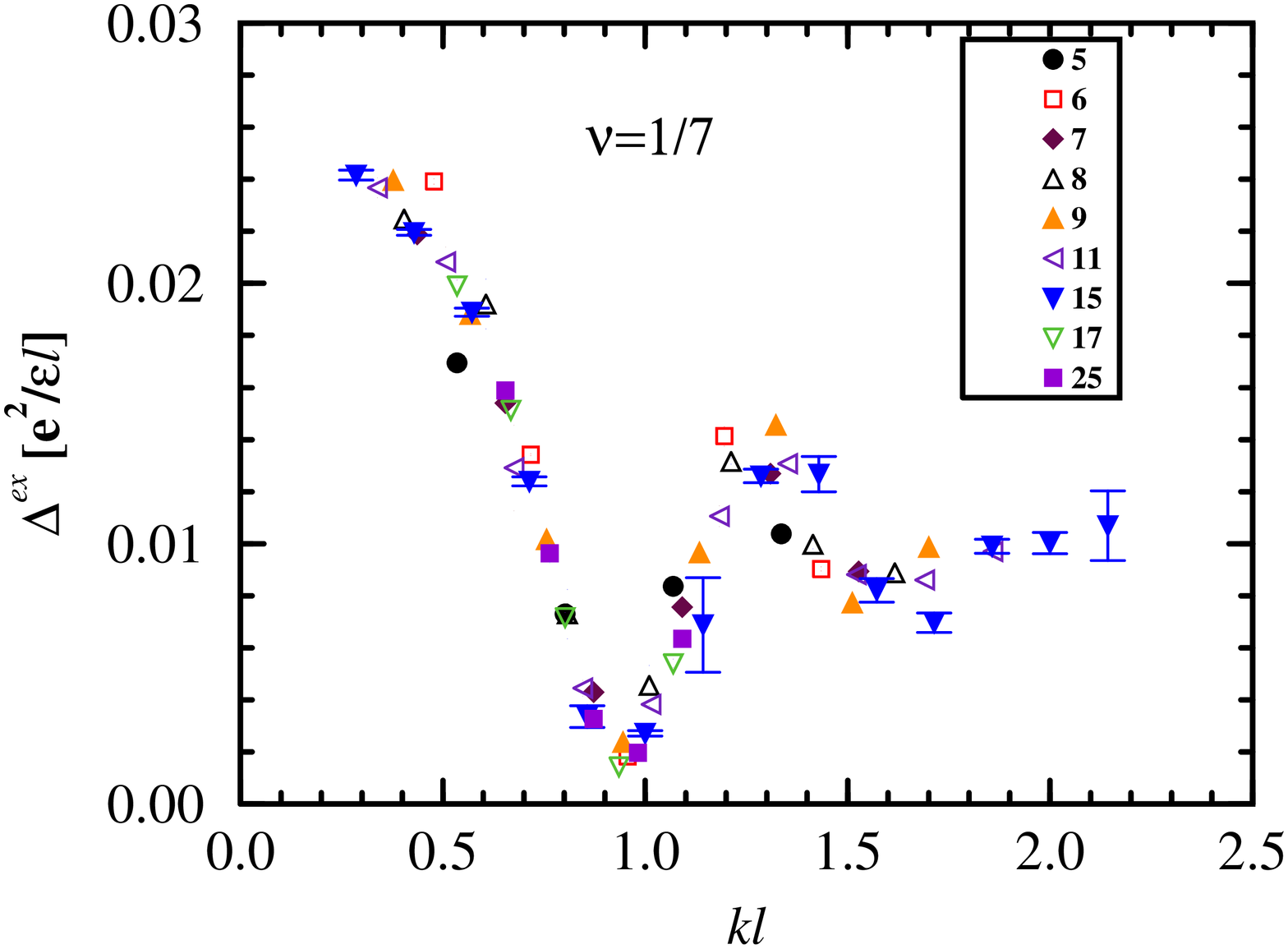,width=6in,angle=-90}}
\caption{Same as in \protect \ref{fig:rev_Fig5} for $\nu=1/7$.}
\label{fig:rev_Fig6}
\end{figure}

\begin{figure}
\centerline{\psfig{file=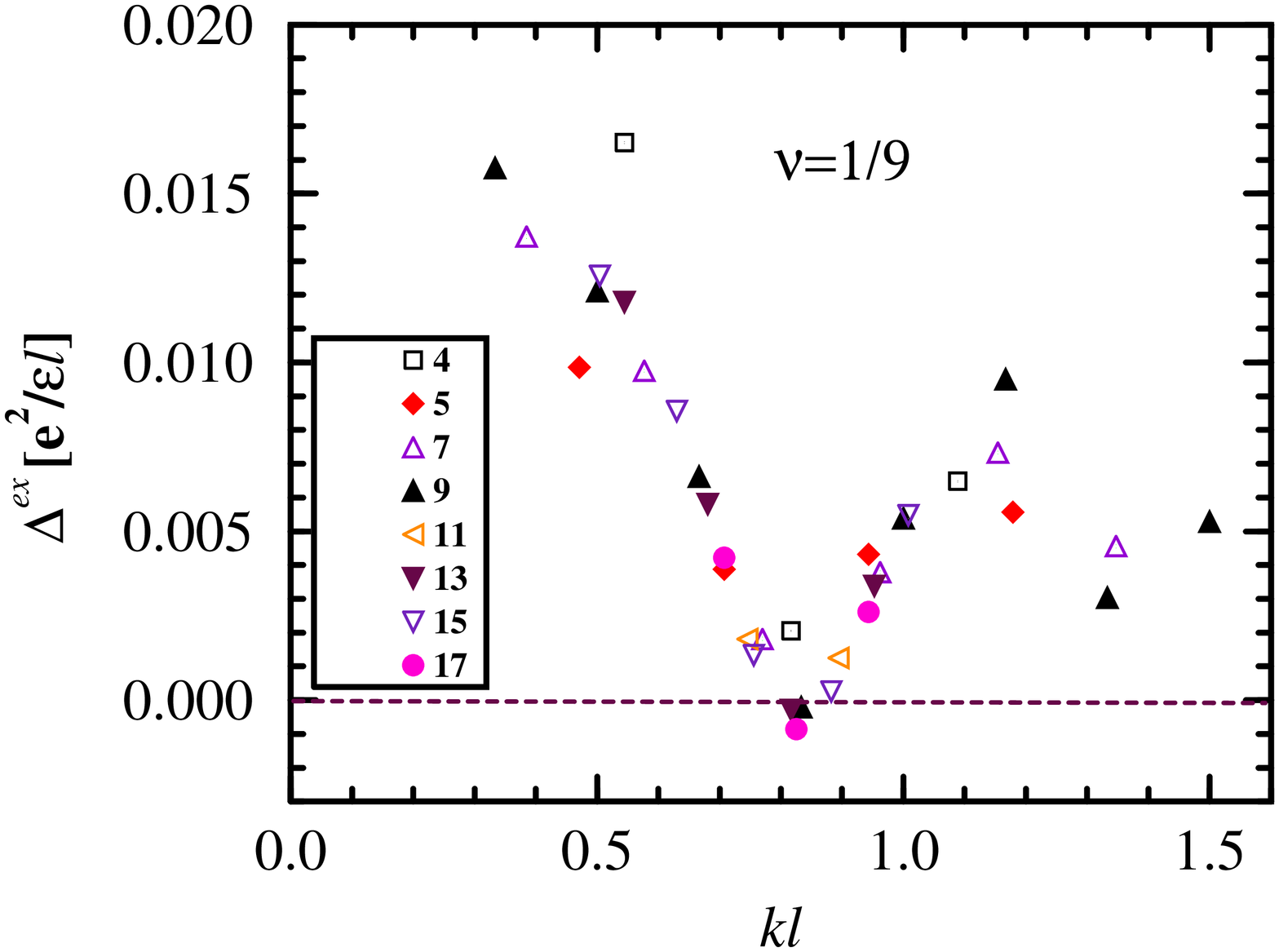,width=6in,angle=-90}}
\caption{Same as in \protect \ref{fig:rev_Fig5} for $\nu=1/9$.}
\label{fig:rev_Fig7}
\end{figure}

Figs. \ref{fig:rev_Fig5}-\ref{fig:rev_Fig7} show the energy of the exciton, $\Delta^{ex}$, at
$\nu=1/5$, 1/7, 1/9 and 2/5 (measured relative to the ground 
state energy)
as a function of the wave vector  $k\equiv L/\sqrt{Q}$. 
We have corrected both the ground and the exciton energies  
for finite size deviation of the density, as discussed earlier.
The fact
that the results for systems with different numbers of electrons fall
on a single line shows that remaining finite size effects are small.
For strictly non-interacting composite fermions, one would expect 
no dispersion of the exciton energy; the dispersion is a measure of  
the interaction between the quasiparticle and the quasihole as 
their relative separation is varied. The oscillatory structure 
arises because of the internal oscillations
in the density of the quasiparticles.

The most striking result is that the energy
of the CF exciton at $\nu=1/9$ falls below that of the ground  
state at approximately $kl\approx 0.85$. The FQHE  
is thus explicitly demonstrated to be
unstable to a spontaneous creation of CF excitons.
Since the size of the minimum energy exciton
($kl^2/\nu\approx 7.5 l$) is
approximately equal to the interparticle separation, a large
number of excitons can be created before their interaction 
may limit their further production. This will clearly 
lead to a collapse of the FQHE, signaling the destruction 
of composite fermions through a vortex unbinding 
transition. [We note that there is another mechanism for
a transition to the Wigner crystal state:  for $\nu \leq 1/72$ the
Laughlin wave function itself describes a Wigner
crystal \protect \cite {Laughlin}. Here the composite fermions are not
destroyed, i.e., the FQHE liquid of composite fermions
freezes into a Wigner crystal of composite fermions.]
Furthermore, the fact that
the wave vector of the instability is
close to the reciprocal lattice vector of the WC \cite {Bonsall} at
$\nu=1/9$, $k_{WC}l=(4\pi \nu/\sqrt{3})^{1/2}\approx 0.898$
also gives a hint to the nature of the true ground state.
Similar result will
undoubtedly be obtained for $\nu<1/9$, ruling out FQHE for
$\nu \leq 1/9$.

Earlier studies \cite {Lam} have compared the energy of the 
Laughlin state with that of the Wigner crystal state, and 
predicted a transition at $\nu^{-1}=6.5\pm 0.5$ (see, Fig. \ref{fig:rev_Fig2} for 
example). In particular,
it has been well established that the incompressible FQHE 
state is not the ground state at 1/9. However, this is the 
first time that an instability of a FQHE
state has been found, despite a number of previous
studies investigating the neutral excitations \cite {Lopez,He}. 
(It should be noted, however, that the field
theoretical approaches of \cite {Lopez} may {\em in principle}
be capable of ruling out FQHE by finding either instabilities or
other lower energy saddle point solutions.)
It is remarkable 
that this can be discovered entirely within the tightly
constrained, zero-parameter framework of the composite fermion theory.
We note that this instability is also an explicit 
demonstration of the breakdown of the mean-field description 
of the CF state, as a result of level crossings in going from
the mean field description to the true state. 

At $\nu=1/7$, the minimum energy of the exciton, $\Delta^{ex}_{min}$,
becomes rather small but remains positive.
This does not rule out a WC ground state here, since a first order
phase transition to a WC state may take place {\em before} 
$\Delta^{ex}_{min}$ hits zero. Nonetheless, our result indicates that 
the question of which state is the ground
state at $\nu=1/7$ deserves a more careful analysis. 
Comparison with Laughlin wave function favors \cite {Lam} a WC ground
state here, but the Laughlin wave function becomes 
progressively worse as one goes to smaller filling factors 
and may not be sufficiently
accurate to lead to a definitive conclusion at $\nu=1/7$.
To elaborate, in the numerical results of Ref. \cite {Fano}, the 
energy of the Laughlin wave function at $\nu=1/5$ is found to 
deviate by $\approx$ 0.15\% from the exact energy already for 
7 electrons. The error in the thermodynamic limit will be more 
and for $\nu=1/7$ still greater. While the error is quite small 
in absolute terms, it may be significant while comparing with the 
Wigner crystal state. Our accurate estimate for the energy of
the Laughlin 1/7 state is -0.280944(6) $e^2/\epsilon l$ while the
best estimate for the WC energy
is -0.2816(5) $e^2/\epsilon l$ \cite {Lam}. The difference is 
$\sim 0.2 \%$, which is likely to be small compared to the 
amount by which the Laughlin state overestimates the energy of
the FQHE liquid at 1/7 in the thermodynamic limit. On the 
other hand, the energy of the WC 
state can also surely be improved. Thus, further theoretical 
work will be needed to provide a conclusive answer to which 
state has lower energy. It is likely that the issue will eventually
be settled by experiments. In this context, it is worth pointing
out that in experiments in semiconductor heterjunctions \cite 
{Goldman} a significant reduction
of the longitudinal resistance (15-20\%) at $\nu=1/7$ has been 
reported, providing evidence of FQHE at 1/7 (not 
conclusive though, as no plateau has been seen
in the Hall resistance). A scenario resulting in such behavior 
can be constructed \cite {Goldman}
by noting that the liquid state will be the 
ground state only in a small region around 1/7; 
somewhat away from 1/7 the Wigner crystal state will become the
ground state due to a cusp in energy at $\nu=1/7$. Due to 
disorder induced spatial fluctuations 
in $\nu$, the liquid state will therefore be the ground state 
only in a region of space where $\nu$ is
sufficiently close to 1/7. Therefore, except for very small 
disorder, the WC
is expected to percolate, but lakes of the incompressible liquid
states should provide a reduction of the longitudinal resistance. At
sufficiently small disorder, the regular FQHE behavior with
well established plateaus may be expected.

\begin{figure}
\centerline{\psfig{file=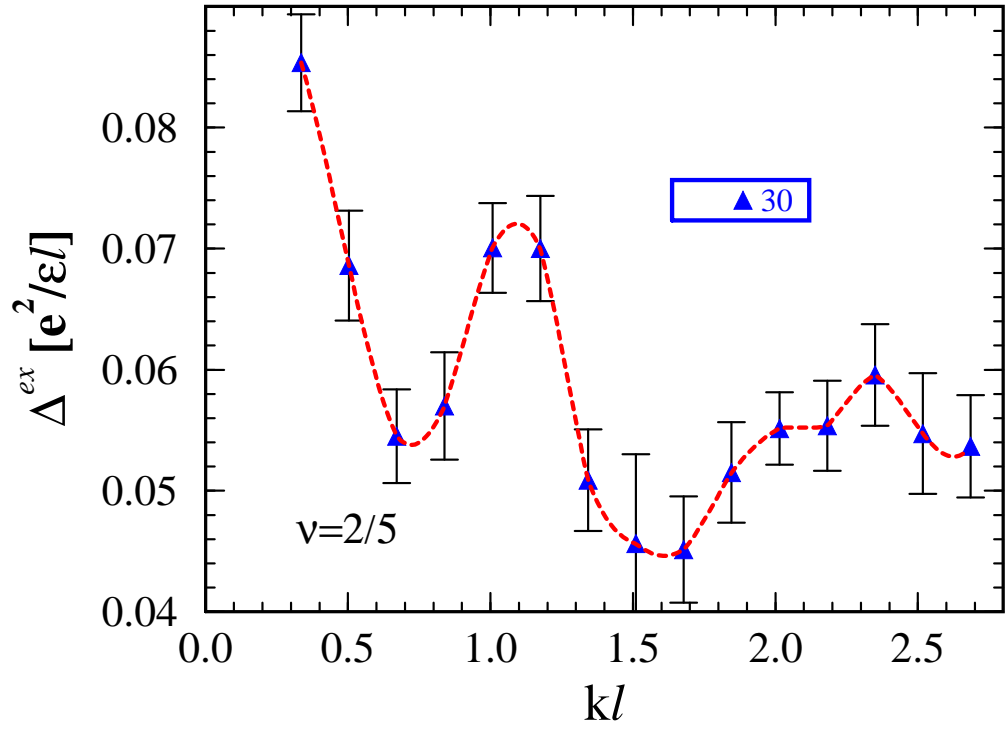,width=6in,angle=-90}}
\caption{ Exciton dispersions for  $\nu=$ 2/5 for $N=30$. The dashed
line is a guide to the eye. }
\label{fig:rev_Fig8}
\end{figure}

Finally, we come to the 
exciton dispersion at 2/5. The CF wave functions are more 
complicated here, so only $2\times 10^7$ the Monte Carlo have been 
performed for each energy, hence the larger statistical error 
in $\Delta^{ex}$ in Fig. \ref{fig:rev_Fig8}. There are two strong
minima at $kl\approx 1.5$ and $kl\approx 0.75$,
consistent with our conjecture in Ref. \cite {Kamilla},
based on the analogy  to the IQHE, that the dispersion of the
exciton of the $n/(2pn+1)$ FQHE state has $n$ principal minima.
We emphasize that even
the biggest spherical system studied previously (with $N_e=10$ particles
\cite {He}) did not produce either the second minimum or the rise
in the energy at small $k$. In general, larger systems are required
for a reasonable description of the large-$n$ FQHE states,
since the relevant number of particles here is the
number of composite fermions in the topmost CF-LL,
which is (approximately) only $N_e/n$. $\Delta^{ex}$ can
similarly be computed for other FQHE states.  We reiterate that no
instability to a WC state is expected as one approaches $\nu=1/2$
along the sequence $n/(2n+1)$, since the energy
of the incompressible CF state remains substantially lower than
that of the WC (Fig. \ref{fig:rev_Fig2}).

Aside from the {\em qualitatively} new physics of instability,
our study provides improved {\em quantitative}
estimates for the excitation energies. The Table V quotes  
the thermodynamic estimate for the 
minimum energy required to create a CF exciton,
$\Delta^{ex}_{min}$, for several FQHE states, 
along with the predictions using the single mode approximation.
(Since both approaches use the same ground state -- 
the Laughlin wave function -- the results indicate  
a lower energy for the CF exciton wave function compared to the
SMA wave function.) The exciton energy is of experimental 
significance. The  Raman scattering experiments
are capable of directly measuring the energy of the exciton
at the minima (or maxima) of the dispersion due
to a peak in the density of states here \cite {Pinczuk}
(although some disorder is required to provide a coupling to 
these modes, otherwise forbidden due to wave vector conservation).
In addition, being the lowest energy excitation,
$\Delta^{ex}_{min}$ is a relevant parameter for
a number of other issues, such as the specific heat,
phonon absorption \cite {Mellor}, and
the finite temperature competition between the
liquid and the WC phases \cite {PP}.

As noted in \cite {GMP}, the dispersion curves in Figs. \ref{fig:rev_Fig5}-\ref{fig:rev_Fig8}
become questionable at small $kl$, where a state with 
two excitons, each with energy $\Delta^{ex}_{min}$, will have
lower energy. There is some question as to whether the Raman peak
associated with the $kl\approx 0$ excitation of the 1/3 state
\cite {Pinczuk} is a single-exciton mode or a two-exciton (bound)
state \cite {Platzman,AMorf}, since the  energy of the single exciton at
$kl\approx 0$ is close to $2\Delta^{ex}_{min}$ \cite {GMP}.
At $\nu=1/5$, on the other hand,  the energy of the single-exciton
mode at small $kl$ is approximately $5\Delta^{ex}_{min}$, which
should make it easier to distinguish it from a two-exciton state.

In summary for this section, we have demonstrated that the CF 
scheme is now demonstrably capable of making detailed quantitative 
predictions, and, in particular, has sufficient accuracy 
to explain the collapse of the FQHE at small $\nu$, which
manifests itself through a finite wave vector excitonic instability.

\section{Composite fermions without higher Landau levels}

As discussed above, the complex system of strongly correlated 
liquid of electrons in the lowest LL can be understood simply in terms  
weakly interacting composite fermions moving in a reduced 
effective magnetic field and occupying, in general, several
Landau levels of their own. Most of the essential features
of experiments can be explained from this very simple 
picture, so it is no surprise that 
the microscopic formulations of composite fermions 
\cite {Jain,HLR,Lopez} have closely followed this physical picture.
In particular, the wave function of composite fermions 
is obtained by taking the wave function of 
non-interacting electrons in the reduced magnetic field,
occupying several LLs in general, 
attaching an even number of vortices to each electron to 
convert it into a composite fermion, and projecting the 
resulting wave function on to the lowest LL. Our confidence 
in this approach, referred to as the ``standard approach" below,
is further bolstered by the excellent quantitative 
agreement between these wave functions and the exact 
eigenstates.

Successful as this theory is, one cannot help but feel intrigued 
by the paradoxical feature that the analogy to higher LLs seems to 
play a central role in explaining the physics of electrons 
confined largely to the lowest LL.  To be sure, there is nothing 
conceptually wrong 
with that. FQHE can and does occur at finite magnetic
fields, when electrons are {\em not} fully confined to the lowest LL.
And even in the limit of very high 
magnetic fields, when the electrons are strictly in their  
lowest LLs, all that one has done is travel somewhat 
outside of the allowed Hilbert space to a point where 
the relevant physical correlations are manifest, lock in this 
physics, and then continue back to the original Hibert 
space hoping on the way that it is not lost. 
In fact, analogy 
to higher LLs is one of the great strengths of the composite fermion
theory since it facilitates a unified description 
of the IQHE and the FQHE and allows a back-of-the-envelope 
understanding of the key experimental facts without 
recourse to any  microscopic theory whatsoever.

Nonetheless, it would be desirable to understand 
composite fermions within the lowest LL. 
The challenge is not to write the CF wave functions in the 
lowest LL -- the standard CF theory already accomplishes this -- but
to understand how to obtain within 
the lowest LL composite fermions along with
their effective field, their effective LL structure,
and effective kinetic energy, concepts 
that appear at the moment to be necessarily tied to  
the analogy with higher LLs. LLL composite-fermion wave functions 
have also been obtained in Ref. \cite {Ginocchio} from a 
different perspective, which does not 
capture the intuitive physics associated with composite fermions
in an obvious manner.

The LLL form of the CF wave functions that we described above
sheds light on this issue. In order to emphasize 
this point, we will now show how the CF theory can be made plausible
entirely within the lowest LL, i.e., develop what will be termed
below a ``LLL approach" for the CF theory, and answer the following  
questions: What is the structure of CF wave 
function in the lowest LL and how does it originate 
from the repulsive interelectron interactions? How does it  
produce an effective magnetic field? 
What is the meaning of the effective cyclotron energy and  
higher-LL-like physics in the lowest LL?
In the end, we will show that the LLL approach is 
identical to the standard one.

For this purpose, it is more convenient to work in the disk 
geometry (i.e., the planar geometry with circular gauge),
to which we will now switch. This will also serve as an extension 
of our new projection approach to the disk geometry, the consideration
of which is useful in certain situations, e.g., for 
two-dimensional quantum dots in high magnetic fields. 

\subsection{Basis states for a single electron}

In the circular gauge, the 
single particle eigenstates of an electron are given by
\begin{equation}
\eta_{n,m}=N_{n,m} e^{-\frac{1}{4}|z|^2} z^m L^m_n(\frac{zz^*}{2})\;,
\label{disk}
\end{equation}
where $z=x-iy$ denotes the position of the electron, 
the magnetic length is set equal to unity,
the normalization coefficient
\begin{equation}
N_{n,m}=\sqrt{\frac{n!}{2\pi 2^m (n+m)!}}\;\;\;,
\end{equation}
$n=0,$ 1, 2,... is the LL index, and the angular momentum is given
by $m=-n, \; -n+1, \; ...$ In the lowest LL ($n=0$), 
the single particles basis states have a particularly simple form,
given by (omitting normalization factors)
\begin{equation}
\eta_{0,m}(z)= e^{-\frac{1}{4}|z|^2}z^{m}\;.
\end{equation}

\subsection{Basis states for a single composite fermion}

\begin{figure}
\centerline{\psfig{file=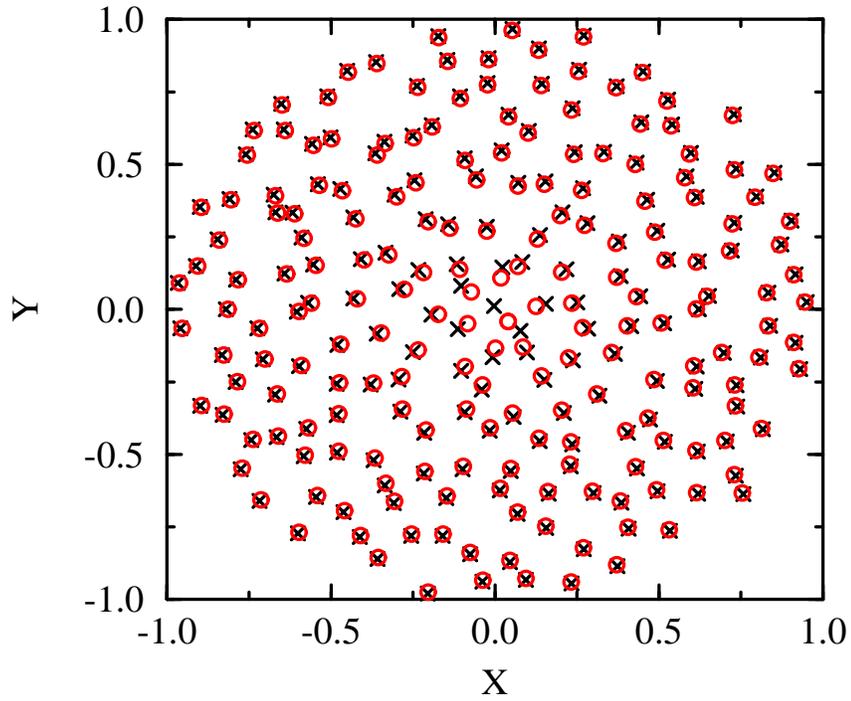,width=6in,angle=-90}}
\caption{Positions of zeros (dots) of $F_1[z,\{z_j\}]$
for a given configuration of  $z_j$ (crosses)
for $N=200$. The other electrons ($z_j$) are taken to be in a
reasonably probable configuration in the presence of a repulsive
interaction.  A defect is seen near the origin. }
\label{fig:rev_Fig9}
\end{figure}

Now, let us consider an (unsubscripted) electron  at $z$ in 
the presence of $N$ other electrons at $z_j$ (for a total 
of $N+1$ electrons), which, at the moment,
are treated simply as repulsive scatterers, and ask how 
the above basis states may be modified.
Our aim is to write correlated basis states of the form
\begin{equation}
e^{-\frac{1}{4}|z|^2}z^{m} F[z,\{z_j\}]\;,
\end{equation}
where the function $F$ incorporates the effect of a repulsion 
between $z$ and the other electrons $z_j$.
The only way our electron can avoid
the $j$th electron is through a factor like $(z-z_j)$, i.e., 
by creating a zero at $z_j$. (Factors
of the type $|z-z_j|^{\alpha}$, which are nonanalytic in $z$,
are not allowed due to the lowest LL restriction.) The first 
guess would be  
\begin{equation}
F_0[z,\{z_j\}]=\prod_j (z-z_j)
\end{equation} 
which has $N$ zeros of $z$, precisely located at the positions of
other electrons, ensuring that $z$ avoids all electrons. 
However, to deal with the general situation, 
we must find ways of constructing $F[z,\{z_j\}]$ 
with different numbers of zeros of $z$. 
It might seem most straightforward to create 
{\em additional} zeros, but this leaves us with too much 
freedom. We therefore resort to {\em removing} some of the
zeros of $F_0$. First consider removing one of the $N$ zeros.
The zero at $z_k$ can be eliminated by  
writing $(z-z_k)^{-1}\prod_j (z-z_j)$. We symmetrize this function
with respect to $z_k$ and write
\begin{equation}
F_1[z,\{z_j\}]=\left[\sum_{k} (z-z_k)^{-1}\right] \prod_j (z-z_j)\;.
\end{equation} 
This function of $z$ has $N-1$ zeros, which are {\em not} 
at the positions of the particles $z_j$, as can be seen by
noting that one of the $N$ terms does not vanish under the
substitution $z=z_j$. However, since $F_1[z,\{z_j\}]$ {\em almost}
vanishes when $z=z_j$, as $N-1$ of the terms vanish,
it is expected that the zeros of this function are typically close
to the particle positions. We have confirmed this numerically. 
Fig. \ref {fig:rev_Fig9} shows the positions of zeros of $F_1$ (marked by circles) 
for a given configuration of $z_j$ (crosses) for $N=200$. There is a 
deficiency of a single zero near the origin,
but away from this region the zeros are tightly bound to electrons.
Thus, the removal of a zero shows up as a ``defect" in the 
sea of electrons dressed with zeros. 
A study of systems of up to $N=1500$ shows that
the geometric average of the distance between an electron and the 
associated zero is given by 
$\sim 1.43 r_0 N^{-0.477}$, where $r_0$ ($\equiv R/\sqrt{N}$ for the
droplet of radius $R$) is the typical interparticle separation, 
indicating that 
the zeros of $F_1[z]$ localize on to particles in the thermodynamic 
limit.  Generalizing, $n$ zeros are removed, creating $n$  defects, 
by writing  
\begin{equation}
F_n[z,\{z_j\}]=\left[\sum_{\{k_i\}} \prod_{k_i}(z-z_{k_i})^{-1}\right] 
\prod_j (z-z_j)\;,
\end{equation}
where the sum $\{ k_1, k_1, ... k_n\}$ is over all distinct 
$n$-tuplets. Apart from an unimportant overall factor, $F_n$ 
can also be expressed in the following equivalent forms:
\begin{equation}
F_n[z,\{z_j\}]
=\left[\sum_{j} \partial_j\right]^n \prod_j (z-z_j)\;
\end{equation}
\begin{equation}
F_n[z,\{z_j\}]
=\partial^n \prod_j (z-z_j)\;,
\end{equation}
where $\partial\equiv\partial /\partial z$ and $\partial_j\equiv\partial
/ \partial z_j$. 

The form of $F_n[z,\{z_j\}]$ multiplying the single electron 
wave function satisfies, and in fact 
is {\em completely} determined by the following 
requirements:

(i) It is restricted to the lowest LL. 

(ii) It has $N-n$ zeros.

(iii) It is a symmetric function of $z_j$.

(iv) It is a homogeneous function of $z$ and $z_j$, i.e., 
$F_n[a z,\{a z_j\}]=a^{N-n} F_n[z,\{z_j\}]$. In other words, it is 
an eigenstate of the total angular momentum.

(v) It is invariant under a translation of all coordinates by 
the same amount.

The above scheme allows us to construct $F$ with 0 to $N$ zeros.
Situations with more than $N$ zeros can be handled straightforwardly
by writing  
\begin{equation}
F_{p,n}[z,\{z_j\}]=  \prod_j (z-z_k)^{p-1}
\partial^n \prod_{k} (z-z_k) 
\end{equation}
in which $(p-1)N$ zeros are bound exactly to electrons,
and the other $N-n$ zeros are the same as before. 
The zeros are more tightly bound to electrons in this 
function than another possible choice 
$$ \partial^n\prod_{k} (z-z_k)^p\;;$$
however, it is likely that both are valid.

We define the basis functions of a composite fermion as
(with the coordinates of other electrons treated as parameters):
\begin{equation}
\eta_{p,n,m}^{CF}(z) \equiv e^{-\frac{1}{4}|z|^2}
z^{m+n} \prod_j (z-z_k)^{p-1} \partial^n \prod_{k} (z-z_k)\;,
\label{scf}
\end{equation}
with $m=-n, -n+1, ...$.
These are nothing but {\em correlated} basis functions of an electron, 
with some knowledge of repulsion between the electron under 
consideration and the other electrons explicitly built into them.
They clearly are not 
orthogonal, but are in general linearly independent, so 
an orthogonal basis may be constructed. (Indeed, if we average 
over $z_j$,  $\eta_{p,n,m}^{CF}(z)$ and 
$\eta_{p,n',m'}^{CF}(z)$ are orthogonal except for $m=m'$.)

What is the energy of a composite fermion 
in state $\eta_{p,n,m}^{CF}(z)$? 
If we assume that the defects are independent, the 
energy of a composite fermion is proportional to the number of 
defects.  We 
assign an energy of $n E_D$ to a composite fermion with $n$ defects, 
where $E_D$ is the (unspecified) energy of a single defect 
(with appropriate averaging over $z_j$). This will be referred to as the
independent-defect approximation (IDA). We will also see below that the 
simple act of counting the number of defects   
gives a considerable amount of insight into 
the energetics of the problem through the IDA with a minimal 
amount of work. Further, the Coulomb energies of the 
CF wave functions, which can be computed by Monte Carlo, will be
shown to be excellent.

An attractive feature of the single CF wave functions is that
away from the defects the zeros are bound to electrons; i.e., 
there are no ``wasted zeros". The no-wasted-zeros was
an important property of the Laughlin wave functions for the
ground states at LL fillings of $\nu=1/(2p+1)$ \cite {Laughlin},
which, however, could not be extended to the many-electron wave
function at other fractions, even as a matter of
principle. The present arguments shows how it can be restored quite
generally, to an extent, at the level of the single CF wave function. 

Finally, defects at $\zeta_1,\; ..., \zeta_n$ are created by replacing 
$\partial^n$ by $\prod_{l=1}^n(\partial-\zeta_l^*)$ in Eq.~(\ref{scf}). The  
resulting wave function will be a linear superposition 
of $\eta_{p,n,m}^{CF}(z)$. In the explicit examples below, we
will only consider compact states composite fermions
which will have no dependence on the positions of the defects 
after antizymmetrization; i.e.,  the replacement mentioned 
above will leave the full wave function unchanged. 
This is analogous to the wave function
of an integer number fully occupied LLs of electrons: one may  
start with Wannier wave functions for 
electrons localized at $\zeta_j$, but the antisymmetrized 
many-body wave function is independent of 
$\zeta_j$.

\subsection{Effective magnetic field}

One of the central features of the CF theory is the 
concept of effective field or the effective filling factor of 
composite fermions. This is a consequence of the fact that the zeros
in the lowest LL are actually vortices, i.e., have phases associated
with them. Let us take $z$ in a closed loop enclosing a 
large area $A$ and
ask what is the phase associated with this loop. Aside from the 
Aharonov-Bohm phase, $2 \pi A B/\phi_0$,
there is a phase of $-p 2\pi \rho A$ due to 
the fact that $z$ encloses $p \rho A$ vortices in the loop, 
each contributing a phase of $-2\pi$.
(Here we neglect order unity corrections due to the presence of
defects and an incomplete binding of the electrons and zeros. 
To see the minus sign, recall $z=x-iy$.) 
This is only half of the story though -- 
it is important to remember that, in the final theory other electrons 
will also see $p$ vortices at $z$, i.e., there is 
an additional contribution to the phase due to $p$ vortices going around 
$\rho A$ electrons, which is also equal to $-p 2\pi \rho A$. Equating 
the net phase to the Aharonov Bohm phase due to an effective magnetic
field $B^{CF}$, $2 \pi A B^{CF}/\phi_0$, gives 
\begin{equation}
B^{CF}=B-2p \rho \phi_0\;.
\end{equation}
The filling factor of composite fermions, $\nu^{CF}=\rho \phi_0/B^{CF}$,
is related to the electron filling, $\nu=\rho \phi_0/B$, as
\begin{equation}
\nu=\frac{\nu^{CF}}{2p\nu^{CF}+1}\;\;.
\end{equation}

\subsection{Many-composite fermion states}

The basis states for many composite fermions are, as usual, Slater
determinants of the single CF states:
\begin{equation}
\Phi^{CF}
=\left| \begin{array}{ccccc}
           \eta_{1}^{CF}(z_1) & \eta_{1}^{CF}(z_2) & . & . & . \\
           \eta_{2}^{CF}(z_1) & \eta_{2}^{CF}(z_2) & . & . & . \\
           . & . & . &  &  \\
           . & . &  & . &  \\
           . & . &  &  & .
           \end{array} \right|
 \;,
\end{equation}
where the subscript of $\eta_j$ denotes collectively the 
quantum numbers  $n_j$ and $m_j$ 
of the composite fermion state (for a fixed $p$). 
The total angular momentum of this state is given by
\begin{equation}
M=M^{CF}+pN(N-1)\;,
\end{equation}
where $M^{CF}=\sum_{j=1}^{N} m_j$ and 
the last term comes from $F_{p,n}$. The total IDA energy is 
\begin{equation}
E=N_D E_D
\end{equation}
where $N_D=\sum_{j=1}^{N} n_j$ is the total number of defects.

\subsection{Testing the LLL wave functions}

\begin{figure}
\centerline{\psfig{file=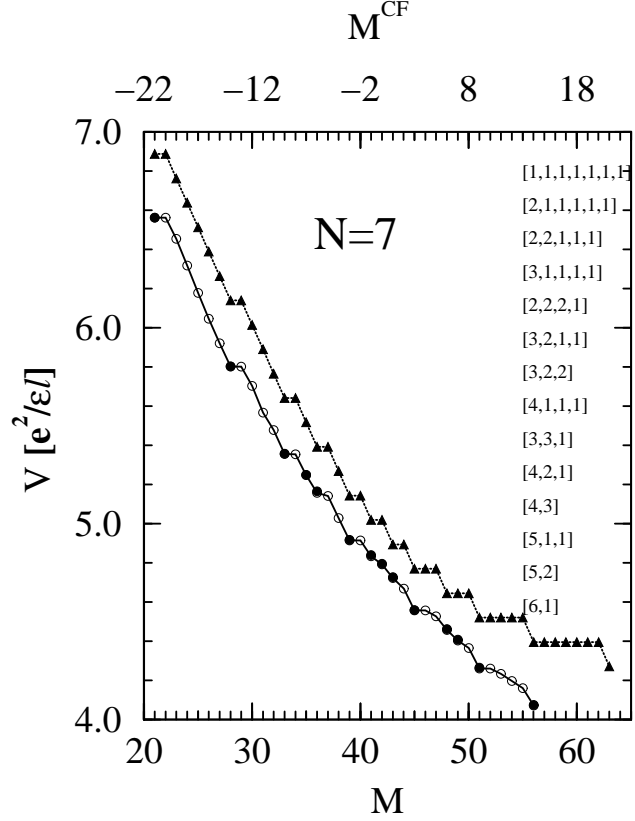,width=4in,angle=0}}
\caption{The empty circles show the exact ground state energy of seven
interacting electrons confined to the lowest LL in the angular
momentum $M$ sector obtained by an exact diagonalization of the
Coulomb Hamiltonian. The filled triangles give the
independent-defect-approximation for the
energy of composite fermions in $M^{CF}$ state, $N_D E_D$,
where $N_D$ is the minimum number of defects at $M^{CF}$ and $E_D$
is the energy of a single defect, chosen to be \protect
$E_D= 0.125 e^2/\epsilon \l$ to obtain the best fit (also,
an arbitrary overall constant has been added to the defect energy).
The filled dots show the Coulomb energies, evaluated by
Monte Carlo. The configurations of the compact CF states are
indicated on the figure  
in sequential order. The wave functions of these 
states contain no adjustable parameters.  The lines are a guide to the eye}
\label{fig:rev_Fig10}
\end{figure}

The many-CF wave functions constructed above are {\em trial} wave 
functions, whose validity remains to be confirmed. 
We again resort to studies of small
systems for which exact solutions can be obtained numerically by 
an exact diagonalization of the Coulomb Hamiltonian in the 
subspace of the lowest Landau level. 
We work in the planar 
geometry with free boundary conditions  -- 
confinement is achieved here 
simply by fixing the angular momentum. 
The exact ground state energy of seven 
electrons interacting via the Coulomb interaction is  
plotted in Fig. \ref {fig:rev_Fig10} as a function of $M$.

We begin by asking what insight can be gained simply by 
counting the total number of defects. It is clear that 
some defects will be necessary  
for angular momenta $M^{CF}<\frac{1}{2}N(N-1)$.
It is straightforward to determine the minimum number of 
defects that $N$ composite fermions must possess in order to 
produce a given  
total angular momentum of $M^{CF}=M-N(N-1)$ 
(we take $p=1$ here). The IDA ground
state energy $N_D E_D$ is also plotted in Fig. \ref {fig:rev_Fig10}, with 
$E_D = 0.125  e^2/\epsilon l$, chosen empirically to get the
best fit. There is a remarkable similarity between the shapes 
of the two curves and the structure of the cusps, 
providing strong support for the qualitative validity of 
the CF picture and our IDA association of the energy with 
the number of defects.

We next carry out a more quantitative test of the microscopic 
structure of the CF wave functions. The number of IDA ground states 
at $M^{CF}$ is tremendously 
small compared to $L$, the total number of electron states at a given
$M$. The most dramatic reduction of the low energy Hilbert space 
due to the formation of composite fermions occurs at those 
values of $N$ and $M$ for which there is a unique IDA ground state.
The CF ground state here has a compact 
pyramid-like structure: the $N_n$ composite fermions with $n$ defects  
occupy the smallest angular momenta (i.e., $m=-n, -n+1, ... -n+N-1$),
and $N_0 \geq N_1 \geq N_2 ...$. 
The wave function of such a  
compact many-CF state is a single Slater determinant, denoted 
by $[N_0, N_1, ...]$. We have computed the Coulomb 
energies of a number of compact CF ground states using Monte Carlo
techniques. As seen in Fig. \ref{fig:rev_Fig10}, the results 
are virtually indistinguishable from the exact ground state
energies. Table V shows the CF energies for a number of compact  
states  for up to seven particles. The 
deviations from the exact energies typically appear
in the fourth significant digit, and are small compared to the
differences between energies at different $M$.  These results 
provide a substantial, non-trivial test of the CF theory
for two reasons.  First, the wave functions of compact CF states   
contain no adjustable parameters; they are completely
determined by symmetry within the space of the CF wave functions.
Second, the size of the {\em many-electron} Hilbert space, $L$,
(which is the dimension of the matrix diagonalized to obtain the
exact ground state) is in general quite large, as shown in 
Table V. (This was incidentally also true for the examples 
discussed earlier in the context of the spherical geometry.)
The only exception is at
$M=N(N-1)/2$, for which there is one one state in the lowest LL, the
$\nu=1$ state. This corresponds to the $[1,1,...]$ of composite
fermions. In this case, the LLL CF wave function is exact for
trivial reasons, as also confirmed explicitly in our calculations.
Note also that when there is more than one compact
state possible for a given $M$ (e.g., for $N=6$
and $M=13$ in Table I), the one with the smallest energy ought to be
identified with the true ground state. 

\subsection{Equivalence between the standard and LLL approaches}

The discussion above makes no reference
to higher LLs, yet obtains the effective magnetic field of
the composite fermions. In this section, we show that 
it is in fact identical to the standard approach. In the latter,
the unprojected wave function of the many CF state, $\Phi^{CF-up}$,
is related to that of the
non-interacting electrons, $\Phi$, as
\begin{equation}
\Phi^{CF-up}=\prod_{j<k}(z_j-z_k)^{2p}\;\Phi\;\;.
\end{equation}
(We note here that the gaussian factor in $\Phi$ is taken
at the external magnetic field. Alternatively, $\Phi$ can be
written at the {\em effective} magnetic field, but then a gaussian 
factor must
be supplied along with the Jastrow factor, which, for bulk FQHE 
states, implies multiplication by $\Phi_1^{2p}$, where $\Phi_1$ is
the wave function of one filled LL \cite {Jain}.) 

For illustration, consider only
states for which $\Phi$ is a single Slater determinant,
made up of single electron states  of
Eq.~(\ref{disk}). Then, 
\begin{equation}
\Phi^{CF-up}= \prod_{j<k}(z_j-z_k)^{2p}
\left| \begin{array}{ccccc}
  \eta_{1}({\bf r}_1) & \eta_{1}({\bf r}_2) & . & . & . \\
  \eta_{2}({\bf r}_1) & \eta_{2}({\bf r}_2) & . & . & . \\
  . & . & . &  &  \\
  . & . &  & . &  \\
  . & . &  &  & .
  \end{array} \right| \;.
\end{equation}
Here the subscript of $\eta_j$ denotes collectively $n_j$ and 
$m_j$, the LL index and the angular momentum, respectively, 
of the electron. 

The total angular momentum of $\Phi^{CF-up}$ is given by
\begin{equation}
M=M^{CF}+pN(N-1)\;\;,
\end{equation}
where $M^{CF}=\sum_j m_j$ is the angular momentum of fermions 
in state $\Phi$. 
In the mean-field approximation (MFA) of the standard 
picture, the composite fermions are treated as non-interacting,
so their total energy is equal to the kinetic energy of electrons 
in the state $\Phi$, with the cyclotron energy replaced by an 
effective cyclotron energy of composite fermions; i.e., 
\begin{equation}
E=[\sum_j n_j] \hbar\omega^{CF}\;.
\end{equation}

The equivalence between
the the LLL and the standard schemes follows if  
we make the following identifications:
\begin{equation}
number\; of\; defects = LL \; index\;,
\end{equation}
anticipated by our notation, and 
\begin{equation}
energy\; of\; a\; defect\; (E_D)= 
effective\; cyclotron\; energy\; (\hbar\omega^{CF}).
\end{equation} 
The IDA of the LLL approach is then identical to the MFA  of the 
standard scheme.  In particular, the curve of the defect 
energy of composite fermions in Fig. \ref{fig:rev_Fig10} is identical to that considered 
earlier in  \cite {Kawamura} for the effective kinetic energy of 
noninteracting fermions at $M^{CF}$.

We now establish the equivalence between the two approaches at the 
level of the microscopic wave functions themselves.  Noting as before 
\begin{equation}
\prod_{j<k} (z_j - z_k)^{2p} =\prod_j
\left[\prod_{k}^{'} (z_j - z_k)^p\right]\;,
\end{equation}
where the prime denotes the condition $j\neq k$,
the wave function $\Phi^{CF-up}$  can be written as
\begin{equation}
\Phi^{CF-up}=
\left| \begin{array}{ccccc}
  \eta^{CF-up}_{1}({\bf r}_1) & \eta^{CF-up}_{1}({\bf r}_2) & . & . & . \\
  \eta^{CF-up}_{2}({\bf r}_1) & \eta^{CF-up}_{2}({\bf r}_2) & . & . & . \\
  . & . & . &  &  \\
  . & . &  & . &  \\
  . & . &  &  & .
  \end{array} \right| \;,
\label{phi'}
\end{equation}
where
\begin{equation}
\eta^{CF-up}_{p,n,m}({\bf r}_j)\equiv \eta_{n,m}({\bf r}_j)
\prod_{k}^{'} (z_j - z_k)^p
\end{equation}
is interpreted as the {\em unprojected} wave function of the
$j$th composite fermion \cite {Jaincomm}.
The CF wave function in the lowest electronic LL is obtained by a
LLL projection of $\Phi^{CF-up}$, which can be accomplished
by projecting $\eta^{CF-up}$ to the lowest LL.

The crux of the argument is that a LLL projection of 
$\eta_{p,n,m}^{CF-up}$ produces precisely (apart from normalization 
factors)  $\eta_{p,n,m}^{CF}$
of Eq.~(\ref{scf}), i.e.,  
\begin{equation}
{\cal P} \eta^{CF-up}_{p,n,m}({\bf r}_j)
\equiv \prod_{i}^{'} (z_j - z_i)^{p-1}
{\cal P} \eta_{n,m}({\bf r}_j)\prod_{k}^{'} (z_j - z_k)=
\eta_{p,n,m}^{CF}({\bf r}_j)\;.
\label{seven}
\end{equation}
This shows that 
${\cal P}\Phi^{CF-up}$ of the standard approach is identical to 
$\Phi^{CF}$ of the LLL approach.

To prove Eq.~(\ref{seven}), we first write 
\begin{equation}
\eta_{n,m}(z)=N_{n,m} e^{-\frac{1}{4}|z|^2} \sum_{k=k_0}^{n} 
(-1)^k {{ n+m \choose n-k}} \frac{1}{2^k k!} z^{*k} z^{k+m} \;,
\end{equation}
where $k_0= max(0,-m)$.
Our goal is to evaluate the projection of the product 
$\eta_{n,m}(z,z^*)\prod_{k} (z - z_k)$. 
As shown in Refs. \cite {Girvin,Jain} the projection 
is obtained by first bringing all $z^*$'s to the left of $z$'s in the 
polynomial part multiplying the gaussian factor and then 
making the replacement 
$z^*\rightarrow  2\partial\;,$
while remembering that the derivatives do not act on the gaussian 
factor. I.e., 
\begin{equation}
{\cal P} \eta_{n,m}({\bf r})\prod_{k} (z - z_k)
=\overline{\eta}_{n,m}({\bf r})\prod_{k} (z - z_k)\;,
\end{equation}
with the operator $\overline{\eta}$ given by  
\begin{equation}
\overline{\eta}_{n,m}({\bf r})=N_{n,m} e^{-\frac{1}{4}|z|^2} 
\sum_{k=0}^{n} (-1)^k
{{ n+m \choose n-k}} \frac{1}{ k!} \partial^{k} z^{k+m} \;.
\end{equation}
Substituting 
$$\partial^k z^{k+m} = \sum_{\alpha=k_0}^{k}
\frac{k!}{\alpha !} {{k+m \choose k-\alpha}}
z^{m+\alpha} \partial^{\alpha}\;\;,$$
and using
$$\sum_{k=k_0}^{n} \sum_{\alpha=k_0}^{k}=
\sum_{\alpha=k_0}^{n} \sum_{k=\alpha}^{n}$$ 
the sum over $k$ is seen to be proportional to 
$(1-1)^{n-\alpha}$ which vanishes unless $\alpha=n$. Thus, only 
the term with $k=\alpha=n$ survives, giving
for the operator $\overline{\eta}$: 
\begin{equation}
\overline{\eta}_{n,m}({\bf r})=
N_{n,m} \frac{(-1)^n}{n!} e^{-\frac{1}{4}|z|^2} 
z^{m+n} \partial^{n} \;.
\end{equation}
Substitution into Eq.~(\ref{seven}) reproduces  
Eq.~(\ref{scf}), completing 
the proof of equivalence of the two formulations 
of the CF theory.

The summary of this section is that 
we have shown that the origin of composite fermions,
their effective magnetic field and LL structure, as well as their
wave functions can all be obtained rather straightforwardly within 
the lowest LL. This complementary description 
gives additional insight into the physics of 
composite fermions, the formal structure of their wave functions, and
the role of the LLL projection operator in the standard 
approach.  

\section{Conclusion}

We have developed an explicit
LLL representation of the CF wave functions 
in which the LLs of composite fermions are 
filled in the standard manner using the single particle basis
states of composite fermions, in complete analogy to the 
problem of non-interacting electrons.
The most important consequence is
that it is now possible to study large CF systems, containing
up to 50 composite fermions. Given how much insight has been 
gained from the study of systems with ten or fewer particles,
this represents a significant setp toward 
a better quantitative understanding of the FQHE.
In particular, certain interesting regions of filling 
factor have become accessible for the first time to quantitative
investigations.
We have obtained estimates for the ground state energies 
and gaps (both to charged and neutral excitations) for a 
number of FQHE states. The charge gaps predicted by the 
CF theory are consistent with the general behavior observed 
experimentally and that anticipated by a mean-field 
theory, although more work will be needed to include
the effects of non-zero transverse width and LL mixing before 
the theoretical
gaps may be compared directly to the experimental ones. An 
important fact discovered in the course of this work is 
that the CF theory is capable of explaining
the absence of FQHE at small filling factors ($\nu \leq 1/9$),
where the FQHE is explicitly shown to be unstable to 
a spontaneous generation of CF-excitons.

The new LLL representation of the CF wave functions also
helps resolve the longstanding question: do we necessarily need to use 
the analogy to higher LLs to explain the physics of electrons
confined to the lowest LL? The answer is in the negative.
We have shown that Laughlin-like
arguments can be used to obtain CF wave functions within the
lowest LL of electrons. Concepts like the CF-LL index, CF cyclotron 
energy, CF filling factor, {\em etc.}, which seem to be 
inextricably tied to the analogy with higher LLs, appear naturally
also in our LLL approach. In particular,  
the CF LL index appears as the number of defects in the single-CF 
wave function and the CF cyclotron energy as the energy of a single
defect.

This work was supported in part by the John Simon 
Guggenheim Foundation, the National 
Science Foundation under grant no. DMR9615005,
and by the MRSEC Program  of the National 
Science Foundation under award number DMR 94-00334.
We thank Tetsuo Kawamura, Daniil Kaplan, and K. Park for 
discussions and help with numerical work.

\section{Appendix}

Here we prove the general correctness of the formula
\begin{eqnarray}
{\cal P}Y_{q,n,m} Y_{q',0,m'}
&=&
K e^{iQ\phi_j} \nonumber \\
&&\sum_{s=0}^{n}(-1)^s {{n \choose s}} {{ 2q+n \choose q+n-m-s}}
\left(\frac{\partial}{\partial u}\right)^{s} u^{Q+M+s}
\left(\frac{\partial}{\partial v}\right)^{n-s} v^{Q-M+n-s}\;,
\label{left}
\end{eqnarray}
where $M=m+m'$ and $Q=q+q'$.
(Here, $K$ represents factors that will not be relevant for
the arguments below.)
The validity of this equation was shown
for $|M|\leq Q$ in Section \ref{cfwf}. In order for it
to be valid generally,
the right hand side must vanish identically for
$|M|>Q$, since no states exist in the lowest LL
with $|M|>Q$. This is what we now show.

First consider the case $M<-Q$. This would be possible
when
$$m=-q-\alpha, \;\;\alpha=1,...n\;,$$
$$m'=-q'+\beta,\;\;\beta=0,1,...$$
Then
$$M=m+m'=-Q-\alpha+\beta\;,$$
and the condition $M<-Q$ implies that $\alpha-\beta \geq 1$.
In fact $\alpha-\beta=1,2...n$.
In this range, the binomial coefficients
impose the following constraints on $s$:
$$ s \leq n$$
$$s\leq q+n-m =2q+n+\alpha$$
$$s\geq 0$$
$$ s \geq -m-q = \alpha$$
As a result, the sum over $s$ goes over
$$s=\alpha, \alpha+1, ... n.$$
Now, consider the derivative $\left(\frac{\partial}
{\partial u}\right)^{s} u^{Q+M+s}.$ The exponent of $u$ can
be written as
$$Q+M+s=s-(\alpha-\beta)\;.$$
It is always less than $s$, since $\alpha-\beta\geq 1$,
and also non-negative,
since the smallest value of $s$ is $\alpha$ ($\geq 1$ by
definition). Therefore,
the derivative with respect to $u$ vanishes:
$$\left(\frac{\partial}
{\partial u}\right)^{s} u^{Q+M+s}=0\;.
$$ It can similarly be shown that for
$M>Q$,
$$\left (\frac{\partial}
{\partial v}\right )^{n-s} v^{Q-M+n-s}=0\;.$$

\begin{table}[t]
\caption{The ground state energies
per particle for several systems at
$\nu= $ 2/5 and 3/7 are obtained from the CF wave functions (third
column) and exact diagonalization (fourth column). The fifth and
sixth columns contain the energies of the charged excitation,
in which one composite fermion from
the south pole in the highest occupied CF-LL is removed and placed
on the north pole in the lowest unoccupied CF-LL.
The seventh and eighth columns give the CF gaps and the exact
gaps.  The energies are in units of \protect $e^2/\epsilon l$,
where $l$ is the magnetic length and $\epsilon$ is the background
dielectric constant, and include interaction with the
uniform positively charged background.
The statistical uncertainty in
the last digit(s) of the CF energy is shown in brackets.
The exact results for $N=10$ and 12
are taken from S. He, S.H. Simon
and B.I. Halperin, Phys. Rev. B {\bf 50}, 1823 (1994).
\label{tab:Tab1}}
\vspace{0.4cm}
\begin{center}
\begin{tabular}{|c|c|c|c|c|c|c|c|} \hline
{\bf $ \nu$ } &  {\bf $N$ }  & \multicolumn{2}{c|}{ \bf ground state}
&
\multicolumn{2}{c|}{\bf excited state} & \multicolumn{2}{c|}{\bf gap}
\\  \cline{2-8}
 &    & CF & exact & CF & exact & CF & exact \\ \hline \hline
$\frac{2}{5}$   &  6   &    -0.50034(4)  &   -0.50040
&   -0.48765(6) &   -0.48789 & 0.07615(61)  &   0.07505 \\ \cline{2-8}
  &  8   &   -0.48022(3) &    -0.48024
&      -0.47144(8)   &    -0.47173 &   0.07021(87) &    0.06809 \\
\cline{2-8}
& 10 & -0.46934(7) &  -0.46945 & -0.46254(5)      & -0.46274
&  0.0681(12)  &   0.06706 \\ \hline
$\frac{3}{7}$ &   9  &    -0.49914(7) &   -0.49918
&  -0.49146(8)  &  -0.49162 &   0.0691(14)  &   0.0681  \\ \cline{2-8}
 & 12  &    -0.48251(5)  &   -0.48264
&    -0.47819(5) &   -0.47826 &     0.0518(12)  &   0.0525  \\ \hline
\hline
\end{tabular}
\end{center}
\end{table}

\begin{table}[t]
\caption{The CF energies for ground states at several
filling factors as a function of $N$.
The charge gaps are also shown in each case; the corrected gap
is obtained by subtracting the interaction energy of two
point-like particles of appropriate charges at the opposite poles
(see text). The Monte Carlo uncertainty for the gaps is the same before and after
correction.
\label{tab:Tab2}}
\vspace{0.4cm}
\begin{center}
\begin{tabular}{|c|c|c|c|c|} \hline
$\nu$ & $N$ &  ground state energy   & gap & corrected gap \\ \hline
\hline
$\frac{2}{5}$ & 4 &      -0.550079(59)& 0.1224 &  0.1340(4)  \\
\cline{2-5}
& 6   &      -0.500339(42)  & 0.0761  &   0.0847(6)  \\  \cline{2-5}
& 8     &    -0.480216(33)  & 0.0702   &  0.0773(9)  \\   \cline{2-5}
& 10   &     -0.469341(66)  & 0.0681   &  0.0742(11)  \\   \cline{2-5}
& 14    &    -0.457866(41)  & 0.0652  &   0.0701(13)  \\  \cline{2-5}
& 16   &     -0.454480(60)  & 0.0650   &  0.0697(15)  \\   \cline{2-5}
& 20   &     -0.449783(42)  & 0.0595  &   0.0637(15)  \\   \cline{2-5}
& 30  &      -0.443884(30)  & 0.0573  &   0.0606(19)  \\   \cline{2-5}
& 40  &      -0.441062(38)  & 0.0597  &   0.0626(29)  \\  \hline
\hline \end{tabular}
\begin{tabular}{|c|c|c|c|c|} \hline
$\nu$ & $N$ &  ground state energy   & gap & corrected gap \\ \hline
\hline
$\frac{3}{7}$ & 9   &      -0.499138(71)  & 0.0691  & 0.0727(14)  \\
\cline{2-5}
& 12  &      -0.482508(49)  & 0.0518  & 0.0548(12)  \\   \cline{2-5}
& 18  &      -0.467681(71)  & 0.0498  & 0.0522(27)  \\   \cline{2-5}
& 21  &      -0.463700(78)  & 0.0493  & 0.0514(29)  \\   \cline{2-5}
& 27  &      -0.458678(58)  & 0.0518  & 0.0537(32)  \\  \cline{2-5}
& 36  &      -0.454368(77)  & 0.0436  & 0.0453(46)  \\ \hline
\hline \end{tabular}
\begin{tabular}{|c|c|c|c|c|} \hline
$\nu$ & $N$ &  ground state energy   & gap & corrected gap \\ \hline
\hline
$\frac{4}{9}$ & 16  &       -0.483701(51)  &  0.0485  &  0.0501(15)
\\
\cline{2-5}
& 20  &       -0.475659(33)  &  0.0424  &  0.0438(15)  \\
\cline{2-5}
& 24  &       -0.470423(36)  &  0.0408  &  0.0421(16)  \\
\cline{2-5}
& 28  &       -0.466874(27)  &  0.0386  &  0.0397(19)  \\
\cline{2-5}
& 32  &       -0.464274(54)  &  0.0394  &  0.0405(33)  \\
\cline{2-5}
& 40  &       -0.460707(46)  &  0.0384  &  0.0394(26)  \\ \hline
\hline \end{tabular}
\begin{tabular}{|c|c|c|c|c|} \hline
$\nu$ & $N$ &  ground state energy   & gap & corrected gap \\ \hline
\hline
$\frac{5}{11}$ & 25  &       -0.477021(55)  &  0.0393  &  0.0450(22)
\\
\cline{2-5}
& 30  &       -0.472327(46)  &  0.0405  &  0.0420(19)  \\
\cline{2-5}
& 35  &       -0.468891(20)  &  0.0330  &  0.0344(24)  \\
\cline{2-5}
& 40  &       -0.466561(31)  &  0.0328  &  0.0341(25)  \\ \hline
\hline \end{tabular}
\vspace{0.4cm}
\begin{tabular}{|c|c|c|c|c|} \hline
$\nu$ & $N$ &  ground state energy   & gap & corrected gap \\ \hline
\hline
$\frac{2}{9}$& 4  &        -0.416946(13)  &  0.0274  &  0.0324(1)  \\
\cline{2-5}
& 8  &        -0.374168(13)  &  0.0196  &  0.0228(2)  \\   \cline{2-5}
& 12  &       -0.362759(15)  &  0.0219  &  0.0244(3)  \\   \cline{2-5}
& 20  &       -0.354339(9)  &   0.0176  &  0.0195(5)  \\   \cline{2-5}
& 26  &       -0.351529(11)  &  0.0161  &  0.0178(6)  \\   \cline{2-5}
& 30  &       -0.350329(13)  &  0.0160  &  0.0175(9)  \\   \cline{2-5}
& 40  &       -0.348373(12)  &  0.0154  &  0.0160(11)  \\
\cline{2-5}
& 44  &       -0.347880(18)  &  0.0177  &  0.0184(14)  \\
\cline{2-5}
& 50  &       -0.347252(14)  &  0.0167  &  0.0173(14)  \\ \hline
\hline \end{tabular}
\begin{tabular}{|c|c|c|c|c|} \hline
$\nu$ & $N$ &  ground state energy   & gap & corrected gap \\ \hline
\hline
$\frac{3}{13}$ & 9   &        -0.382711(6)  &  0.0164  &  0.0172(2)
\\
\cline{2-5}
& 12  &       -0.373119(12)  &  0.0136  &  0.0143(4)  \\   \cline{2-5}
& 15  &       -0.367677(15)  &  0.0144  &  0.0150(4)  \\   \cline{2-5}
& 21  &       -0.361826(12)  &  0.0163  &  0.0168(7)  \\   \cline{2-5}
& 30  &       -0.357607(11)  &  0.0160  &  0.0164(7)  \\   \cline{2-5}
& 36  &      -0.35600(17)    &  0.0146  &  0.0150(13)  \\
\cline{2-5}
& 42  &       -0.354869(11)  &  0.0136  &  0.0139(12)  \\
\cline{2-5}
& 51  &       -0.353697(25)  &  0.0151  &  0.0154(26)  \\ \hline
\hline \end{tabular}
\begin{tabular}{|c|c|c|c|c|} \hline
$\nu$ & $N$ &  ground state energy   & gap & corrected gap \\ \hline
\hline
$\frac{4}{17}$& 16  &       -0.372512(33)  &  0.0128  &  0.0131(10)
\\
\cline{2-5}
& 20  &       -0.367865(29)  &  0.0105  &  0.0107(9)  \\   \cline{2-5}
& 32  &       -0.361353(24)  &  0.0129  &  0.0131(17)  \\
\cline{2-5}
& 36  &       -0.360167(33)  &  0.0134  &  0.0136(24)  \\
\cline{2-5}
& 40   &      -0.359214(28)  &  0.0120  &  0.0122(22)  \\
\cline{2-5}
& 48  &       -0.357811(27)  &  0.0109  &  0.0111(27)  \\ \hline
\hline \end{tabular}
\end{center}
\end{table}

\begin{table}[t]
\caption{The thermodynamic estimates for the ground state energy
per particle and the charge gap
for a number of FQHE states.
\label{tab:Tab3}}
\vspace{0.4cm}
\begin{center}
\begin{tabular}{|c|c|c|} \hline
$\nu$ & ground state energy  &  gap \\ \hline
\hline
1/3 & -0.409828(27)    &  0.106(3) \\ \hline
2/5 & -0.432804(62)    &  0.058(5) \\ \hline
3/7 & -0.442281(62)    &  0.047(4) \\ \hline
4/9 & -0.447442(115)   &  0.035(6) \\ \hline
5/11& -0.450797(175)   &  0.023(10)\\ \hline
1/2 & -0.4653(2) & -  \\ \hline
1/5 & -0.327499(5)     &  0.025(3) \\ \hline
2/9 & -0.342782(35)    &  0.016(2) \\ \hline
3/13& -0.348349(19)    &  0.014(2) \\ \hline
4/17& -0.351189(39)    &  0.011(3) \\ \hline
\hline
\end{tabular}
\end{center}
\end{table}

\begin{table}[t]
\caption{The minimum gap to neutral excitation,
$\Delta^{ex}_{min}$, for several filling factors. The SMA prediction
is also given for comparison.
\label{tab:Tab4}}
\vspace{0.4cm}
\begin{center}
\begin{tabular}{|c|c|c|} \hline
$\nu$ & \multicolumn{2}{c|}{Minimum gap energy}\\ \cline{2-3}
&  CF   & SMA  \\ \hline
\hline
$\frac{1}{3}$ &  0.063(3)  & 0.078 \\ \hline
$\frac{1}{5}$ & 0.0095(6) & 0.017 \\ \hline
$\frac{1}{7}$ & 0.0009(5) & 0.0063 \\ \hline
$\frac{1}{9}$ & $<$ 0 & 0.0032   \\ \hline
$\frac{2}{5}$ & 0.040(4)  &  -   \\ \hline
\hline \end{tabular}
\end{center}
\end{table}

\begin{table}[t]
\caption{Comparison between the energy of the CF wave function
$\Phi^{CF}$ and the exact Coulomb energy for the ground states at
several angular momenta $M$ in the disk geometry.
The energies are in units of $e^2/\epsilon l$.
No positive background has been assumed;
confinement is due to fixing the total angular momentum.
The energies of composite fermions in the compact states
$[N_0, N_1,...]$ (notation explained in text) have
been evaluated by Monte Carlo, with the statistical uncertainty in
the last few digits shown in brackets.
The exact energies have been obtained by a numerical diagonalization
of the Coulomb Hamiltonian in $L$ dimensional Hilbert space.
\label{tab:Tab5}}
\vspace{0.4cm}
\begin{center}
\begin{tabular}{|c|c|c|c|c|c|} \hline
$N$& $M$ & $L$ & CF state & CF energy & Exact energy \\ \hline
\hline
 4&  10 & 5  &  [2,1,1]     &   1.78537(40)     &   1.785088 \\
\cline{2-6}
  &  12 & 9  & [2,2]   &     1.68651(33)  &      1.685182 \\
\cline{2-6}
  &  14 & 15  &  [3,1]       &   1.50222(54)     &   1.500658 \\
\hline
 5&  15 & 7  &  [2,1,1,1]   &   2.91991(69)  &     2.918660 \\
\cline{2-6}
  &  20 & 30  &  [3,1,1]     &   2.53756(35)      &  2.536763 \\
\cline{2-6}
  &  22 & 47  &  [3,2]       &   2.43051(61)       & 2.429793 \\
\cline{2-6}
  &  25 & 84  &  [4,1]       &   2.24831(67)      &  2.247239 \\
\hline
 6&  21 & 11  &  [2,1,1,1,1] &   4.26551(45)  &     4.264391 \\
\cline{2-6}
  &  25 & 35  &  [2,2,1,1]   &   3.92361(51)  &     3.921520 \\
\cline{2-6}
  &  27 & 58  &  [3,1,1,1]   &   3.79609(83)  &     3.793702 \\
\cline{2-6}
  &  33 & 199  &  [4,1,1]     &   3.41877(82)  &     3.418581 \\
\cline{4-5}
  &     &  &  [3,3]       &   3.48196(68)  &              \\
\cline{2-6}
  &  35 & 282  &  [4,2]       &   3.29328(89)  &     3.289422 \\
\cline{2-6}
  &  39 & 532  &  [5,1]       &   3.11094(52)  &     3.110306 \\
\hline
 7&  28 & 15  &  [2,1,1,1,1,1] & 5.80341(48)  &     5.802404 \\
\cline{2-6}
  &  33 & 65  &  [2,2,1,1,1] &   5.35812(63)  &     5.356087 \\
\cline{2-6}
  &  35 & 105  &  [3,1,1,1,1] &   5.25064(49)  &     5.247864 \\
\cline{2-6}
  &  36 & 131   &  [2,2,2,1]   &   5.16513(77)  &     5.158199 \\
\cline{2-6}
  &  39 & 248  &  [3,2,1,1]   &   4.91828(46)  &     4.915675 \\
\cline{2-6}
  &  41 & 364  &  [3,2,2]     &   4.84061(30)  &     4.834253 \\
\cline{2-6}
  &  42 & 436  &  [4,1,1,1]   &   4.79650(39)  &     4.792744 \\
\cline{2-6}
  &  43 & 522  &  [3,3,1]     &   4.72768(58)  &     4.723504 \\
\cline{2-6}
  &  45 & 733  &  [4,2,1]     &   4.55992(45)  &     4.557431 \\
\cline{2-6}
  &  48 & 1175  &  [4,3]       &   4.46259(64)  &     4.457815 \\
\cline{2-6}
  &  49 & 1367  &  [5,1,1]     &   4.40837(75)  &     4.404352 \\
\cline{2-6}
  &  51 & 1824  &  [5,2]       &   4.26554(71)  &     4.261579 \\
\cline{2-6}
  &  56 & 3539  &  [6,1]       & 4.07609(29)    &     4.071991 \\
\hline
\hline \end{tabular}
\end{center}
\end{table}

\end{document}